\documentclass[preprint,12pt]{elsarticle}
\usepackage{fullpage, amsmath, amsthm, amsfonts}
\usepackage{hyperref}
\usepackage{nameref}
\usepackage{lmodern}

\usepackage{algorithm}
\usepackage{algorithmicx,rotating}
\usepackage{color}
\usepackage{algpseudocode}
\usepackage{graphicx}
\usepackage{parskip}
\usepackage{natbib}           % for author year citations \citet \citep
\setcitestyle{aysep={}}
\usepackage{setspace}
\def \s2{\sigma^2}
\def \hs2{\hat{\sigma}^2}
\def \siga2{\sigma_{\alpha}^2}
\def \sige2{\sigma_{\epsilon}^2}
\def \sig2{\sigma^2}

\newcommand{\rT}{\mbox{\tiny{T}}}

\newcommand{\bmE}{\mbox{\boldmath $E$}}

\newcommand{\bQ}{\mbox{\boldmath $Q$}}

\newcommand{\bms}{\mbox{\boldmath $s$}}
\newcommand{\bs}{\mbox{\boldmath $s$}}

\newcommand{\bS}{\mbox{\boldmath $S$}}

\newcommand{\bmu}{\mbox{\boldmath $u$}}
\newcommand{\bu}{\mbox{\boldmath $u$}}

\newcommand{\bw}{\mbox{\boldmath $w$}}

\newcommand{\bx}{\mbox{\boldmath $x$}}

\newcommand{\by}{\mbox{\boldmath $y$}}

\newcommand{\bz}{\mbox{\boldmath $z$}}

\newcommand{\beq}{\begin{equation}}       
\newcommand{\eeq}{\end{equation}}       
\newcommand{\beqn}{\begin{eqnarray}}
\newcommand{\eeqn}{\end{eqnarray}}

\newcommand{\bbeta}{\mbox{\boldmath $\beta$}}

\newcommand{\btheta}{\mbox{\boldmath $\theta$}}

\newcommand{\bmphi}{\mbox{\boldmath $\phi$}}
\newcommand{\bphi}{\mbox{\boldmath $\phi$}}

\newcommand{\bmmu}{\mbox{\boldmath $\mu$}}

\newcommand{\bmepsilon}{\mbox{\boldmath $\epsilon$}}

\newcommand{\bSigma}{\mbox{\boldmath $\Sigma$}}

\def\bSig\mathbf{\Sigma}

\usepackage{titlesec}

\begin{document}
\begin{frontmatter}
\title{Estimation of Health and Demographic Indicators with Incomplete Geographic Information}
\author{Katie Wilson$^1$ and Jon Wakefield$^{2,3}$\\
\small{$^1$ Department of Health Metrics Sciences, University of Washington} \\
\small{$^2$ Department of Biostatistics, University of Washington} \\
\small{$^3$ Department of Statistics, University of Washington}}
\date{}

%\begin{document}
%\maketitle

%\newpage
\begin{abstract}
In low and middle income countries, household surveys are a valuable source of information for a range of health and demographic indicators. Increasingly, subnational estimates are required for targeting interventions and evaluating progress towards targets. In the majority of cases, stratified cluster sampling is used, with clusters corresponding to enumeration areas. The reported geographical information varies. A common procedure, to preserve confidentiality, is to give a jittered location with the true centroid of the cluster is displaced under a known algorithm. An alternative situation, which was used for older surveys in particular, is to report the geographical region within the cluster lies. In this paper, we describe a spatial hierarchical model in which we account for inaccuracies in the cluster locations. The computational algorithm we develop is fast and avoids the heavy computation of a pure MCMC approach. We illustrate by simulation the benefits of the model, over naive alternatives.\end{abstract}

\end{frontmatter}
\section{Introduction}
Effective implementation of health programs requires information on where unmet need exists within countries. In many low and middle income countries (LMIC), health data can come from a variety of sources, and in many cases the sources are an incomplete representation of all of the country's inhabitants. Surveys are a primary tool for obtaining vital information. One example of surveys that are commonly used, especially in LMIC, are the Demographic and Health Surveys (DHS). Typically, the DHS employs stratified, cluster sampling, where the clusters represent a small area of the country. Within clusters, households are randomly selected. They are designed to provide reliable estimates at a pre-specified, and usually geographically large, administrative level. However, policymakers and researchers are often interested in modeling and understanding health indicators at lower levels, e.g.,~at the district level.
% or on a 10km $\times$  10km grid.

To protect respondent confidentiality, survey data from households within the same cluster are aggregated to a single point, the centroid of the cluster. This cluster location information can be used for spatial modeling. However, the geographic identifiers available in the DHS can vary and typically the precise coordinates of the cluster centers are not publicly available \citep{burgert:13}. Recently, the geographic locations of the clusters (i.e., the centroids) are provided, but they are displaced. Specifically, urban clusters locations are displaced up to 2km, and 99\% of rural cluster locations are displaced up to 5km, with the remaining 1\% displaced up to 10km. For example, in Kenya displaced GPS data are available for DHS completed on or after 2003. In older DHS and other surveys such as the Multiple Indicator Clusters Surveys (MICS) only the larger administrative area within which the cluster resides is reported. We will refer to this procedure as ``masking''. In the literature this is sometimes referred to ``aggregation'', though not to be confused with the aggregation procedure that was the focus of \cite{wilson:wakefield:20} in which the observed data consist of the sum or mean response of all responses in the area. In that paper, censuses were considered, which provided outcomes that are aggregated over an entire administrative area. Here, we consider point data, but where the geographic location of the point is assigned to the administrative area within which the point belongs. For the purposes of this paper, we will ignore issues involving the first step of aggregating household data to a single point, though an approach similar to that in \cite{wilson:wakefield:20}  could be used. Instead, we focus on issues surrounding displacement and masking of the cluster locations.

First, we consider the displacement scenario. A naive analysis would ignore the jittering of the cluster centroids and fit a continuous spatial model using the displaced location. The effect of doing this in spatial analyses has been studied  \cite{gething:etal:15}. Using real data,  100 datasets were simulated  with jittered location information and  the impact on analyses involving several indicators of interest was assessed. The effect of the displacement on spatial correlation, spatial covariate associations, and model derived surfaces was investigated. In the example, using empirical variograms, it was found that there was not a large impact on spatial correlation. However, some differences in the relationship between spatial covariates and the outcomes was found; models naively using the spatial covariate value at the displaced value tended to have lower $R^2$ although this was not always the case. Some inaccuracies in predicted surfaces were also observed when using displaced data, and these differences tended to be exacerbated when the spatial covariate changed quickly in space. Based on work by ~\cite{perez:13}, the DHS have proposed guidelines that when using spatial covariate raster data, the average value of the covariate in cells within a specified buffer (10km for rural clusters and 2km for urban clusters) of the reported cluster location should be used in analyses. Warren et al.~\cite{warren:etal:16} examine the impact of jittering on covariate modeling, when the covariates are available for areas, and the outcome at points. They introduce a new method for this scenario, maximum probability covariate (MPC) selection and show superior performance with the naive method of using the covariate associated with the displaced point. This method cannot resolve the problem completely, since bias in association parameters will in general result unless the correct covariate is determined with probability 1.

 Having location information subject to positional error can be thought of as an error-in-variables problem. In particular, let $\{\bms_1,\dots,\bms_n\}$ denote the set of true, unobserved,  (analogous to covariates in the regular setting) that give rise to the set of observed outcomes $\{y_1,\dots,y_n\}$. Denote the measured (reported) locations associated with the outcomes as $\{\bmu_1,\dots,\bmu_n\}$. The Berkson measurement error model would be, $\bms_i = \bmu_i + \bmepsilon^\star_i$ where $\bmepsilon^\star_i$ is an error term, whereas the classical measurement error model would be, $\bmu_i = \bms_i + \bmepsilon_i$ where $\bmepsilon_i$ is an error term \citep{carroll:06}. 
 The Berkson model is particularly appealing when there are a set of desired locations that outcomes should be collected at, but the actual location has been perturbed. This could be a result of using an imprecise positional instrument. On the other hand, the classical measurement error model would arise when outcomes are collected at a particular location, but the reported location has been perturbed. Many proposed approaches seeking to address the positional error issue  focus on normally-distributed outcomes under a Berkson measurement error model \citep{gabrosek:02,cressie:kornak:03,fanshawe:diggle:11}. To overcome the computationally expensive Monte Carlo integration from earlier work using a Berkson measurement error model, an approximate composite likelihood for inference has been suggested  \cite{fronterre:etal:18}. In their application to DHS data from Senegal, the displacement mechanism is approximated and the stratified sampling nature of the data is ignored. In this paper, we develop a method under the classical measurement error model and the outcome distribution can be non-normal. 
 %\cite{fanshawe:diggle:11} propose a model for normally distributed outcome data subject to geographic displacement. To relate the true, unobserved location and reported location they view the true location as a function of the reported location and thus use a Berkson measurement error model, rather than a classical measurement error model \citep{carroll:06}. The Berkson model arises when there are a set of intended locations (reported), but the actual observations occur elsewhere. \cite{gabrosek:02,cressie:kornak:03} also study this particular issue and suggest a krigging-based approach.
 %As will be described in the following sections, we argue that a classical measurement error model is more natural for the application we consider. This is because there are a set of true (not reported) locations which gives rise to the data, but the reported location are elsewhere. To evaluate the likelihood, \cite{fanshawe:diggle:11} use Monte Carlo integration. \cite{fronterre:etal:18} build on this work and use an approximate composite likelihood for computational purposes. In the application to DHS data, the displacement mechanism is approximated and the stratified sampling nature of the data is ignored.
 %The discussants of this work, provide further references to this problem, in an environmental sciences context\cite{crainiceanu:11,gelfand:banerjee:11,cressie:11,lindgren:etal:11jitter,fanshawe:diggle:11rej}. Zandbergen 2014 summary

Now, we turn to the issue of masking. To incorporate masked data reported at the administrative area, one solution is to use a discrete spatial model, such as the ICAR model. Using this type of model would not allow for higher spatial resolution maps than the broadest administrative level reported.
%, and more recent surveys provide (jittered) geographic coordinates. 
Further issues could arise if the divisions of regions change over time. Additionally, these boundaries are often arbitrary and using a discrete model can be difficult to interpret if regions differ substantially in size and shape. In the context of modeling the under-five mortality rate, ~\cite{golding:etal:17} fit a continuous spatial model and develop an approximate strategy for including data associated with areas and do not distinguish between aggregate and point-level data with missing coordinates. To deal with the masking problem, points are randomly generated in an area according to the population density. Points nearby are grouped together to form ``pseudo-clusters'' and assigned a weight based on the population that each ``pseudo-cluster'' represents. These weights then essentially partition the observed data to each of the ``pseudo-clusters.'' This approach has no formal justification, and it is difficult to gauge how the method will perform in practice.% This problem, and the aggregate data problem studied in Chapter \ref{chap:pointless} are closely related to ecological inference \citep{wakefield:04read} and the change of support problem \citep{gotway:young:02,bradley:etal:16ACS}, also known as the modifiable areal unit problem \citep{wong:09}.
%As in Chapter \ref{chap:pointless}, we assume that there is a latent, continuous GRF that varies in space, $\{S(s): s \in D \subset \mathbb{R}^2\}$ where $D$ is the study region of interest.

The organization of this paper is as follows.  In Sections \ref{sec:meth}, we make explicit the problem and propose a model that can accommodate masked (only administrative area available) or displaced (jittered coordinates)  data. In Section \ref{sec:comp} a hybrid computational scheme is described. In Sections \ref{sec:sim} and \ref{sec:simres}, we conduct a simulation study to assess the impact of each of these problems on spatial modeling. We also consider disclosure risk, which, in this case, refers to the ability to identify the true cluster location from the reported cluster location. Finally, we conclude with a discussion in Section \ref{sec:disc}, which includes directions for future work.

%NOTE: look into regression calibration? Carroll and Stefanski 1990

\section{Method}
\label{sec:meth}

We suppose that the cluster data is associated with a true point location, namely the centroid of the cluster. Let $i=1,\dots,I$ index the administrative areas, $j=1,\dots,J$ index the strata (typically $J=2$ for urban/rural), and $k=1,\dots, K_{ij}$ index the clusters within administrative area $i$ and strata $j$. 
Consider cluster $k$ in strata $j$ and administrative area $i$, and denote %the outcome data by $y_{ijk}$, 
the true cluster centroid location by $\bms_{ijk}$,
%, (non-spatial) covariate information by $\bx_{ijk}$, spatial covariates depending on the the true cluster location by $\bz_{ijk} = \bz(\bs_{ijk})$, 
and the available location information by $\bmu_{ijk}$. Suppose the set of all possible (true) cluster locations (i.e.,~the sampling frame) is known and denote the set of the potential locations in area $i$, strata $j$ by $\bmE_{ij} = \{\bmE_{ije}, e=1,\dots,m_{ij}\}$.

In the \textit{masking scenario}, only the area in which the cluster is located is reported, which we will denote by $\bmu_{ijk} = \{\bms_{ijk} \in \bmE_{ij}\}$. Hence, the prior on the location is,
\begin{align}
p(\bms_{ijk} = \bmE_{ije} | \bmu_{ijk}) = d_{ije}, \quad e=1,\dots,m_{ij}, \label{eq:maskprior}
\end{align}
where $d_{ije}$ is the probability that potential location $\bmE_{ije}$ was selected. If probability proportional to size (PPS) sampling was undertaken (the usual strategy in the DHS), then $d_{ije} \propto N_{ije}$ where $N_{ije}$ is the population size of enumeration area located at $\bmE_{ije}$. If random sampling was undertaken, then $d_{ije} \propto 1$.

In the \textit{displacement scenario}, a jittered version of the true location is reported, which we will denote by $\bmu_{ijk} = \bms_{ijk} + \bmepsilon_{ijk}$ where $\bmepsilon_{ijk}$ is the result of the jittering probability density function. % In the measurement error literature, this can be thought of as a classical measurement error model. 
We consider the DHS jittering algorithm in which the true location is randomly displaced according to the distribution (in polar coordinates), $p(r,\theta) = (2\pi R)^{-1}I({0 < r<R})\times I({0<\theta<2\pi})$ where $R=2$km for urban clusters and $R=5$km for 99\% of rural clusters and $R=10$km for the remaining 1\% of rural clusters, and $I(\cdot)$ is the indicator function. That is,
$$p(\bms_{ijk} = \bmE_{ije} | \bmu_{ijk}) \propto p(\bmu_{ijk} | \bms_{ijk} = \bmE_{ije}) \times p(\bms_{ijk} = \bmE_{ije}), \quad e=1,\dots,m_{ij},$$
where $p(\bms_{ijk})$ corresponds to (\ref{eq:maskprior}). To derive the first term on the right side, we need to marginalize over possible values of $R$. First note that for a given $R$, 
$$p(\bmu_{ijk}  | \bms_{ijk} = \bmE_{ije}, R) = [2\pi Rd(\bmu_{ijk}, \bmE_{ije})]^{-1} C_{ije,R} I({0 < d(\bmu_{ijk}, \bmE_{ije}) < R})$$
where $d(\bmu_{ijk}, \bmE_{ije} )= [(E_{ije1} - u_{ijk1})^2 + (E_{ije2} - u_{ijk2})^2]^{1/2}$ is the distance between the candidate location
$\bmE_{ije} = [E_{ije1}, E_{ije2}]$ and the reported location $\bmu_{ijk} = [u_{ijk1}, u_{ijk2}]$ and 
\begin{align}
    C_{ije,R} = \left[\int_{u \in D_i} [2\pi R d(\bmu, \bmE_{ije})]^{-1}I({0<d(\bmu, \bmE_{ije}) < R}) ~d\bmu \right]^{-1} \label{eq:ulocs:norm}
\end{align}
is the normalizing constant that accounts for the jittered point being restrictedwil to stay within administrative area $D_i$. Therefore,
\begin{align}
    p(\bms_{ijk} = \bmE_{ije} | \bmu_{ijk}) \propto \begin{cases}
    d_{ije} [4\pi d(\bmu_{ijk}, \bmE_{ije})]^{-1}C_{ije,R=2}I({0 < d(\bmu_{ijk}, \bmE_{ije}) < 2km}) & \text{if urban}\\
    d_{ije} \{ 0.99 \times [10\pi d(\bmu_{ijk}, \bmE_{ije})]^{-1}C_{ije,R=5}I({0 < d(\bmu_{ijk}, \bmE_{ije}) < 5km}) & \\
    + 0.01 \times [20\pi d(\bmu_{ijk}, \bmE_{ije})]^{-1}C_{ije,R=10}I({0 < d(\bmu_{ijk}, \bmE_{ije}) < 10km})\}  & \text{if rural} \label{eqn:jitterprior}
    \end{cases}
\end{align}
where $d_{ije}$ are the prior probabilities on the locations.

Denote the outcome data measured at each cluster as $\by$, the available location information as $\bu$, the true (unobserved) location information as $\bs$, non-spatial covariates as $\bx$, spatial covariates as $\bz$ where $\bz_{ijk} = \bz(\bms_{ijk})$ is the vector of spatial covariates associated with the true location of cluster $k$ in strata $j$ and administrative area $i$.
In contrast to Gaussian Markov Random Fields (GMRFs) that are fundamentally discrete, Gaussian Random Fields (GRFs) are continuously indexed. Consider a domain $\mathcal{D} \in \mathbb{R}^2$. Then $S(\bs)$ is a GRF if all finite collections are jointly multivariate normal. That is, for a collection of points $[\bms_1,\bms_2,\dots,\bms_n]^{\rT}$, the density is,
$$\pi(\bS) = (2\pi)^{-n/2}|\bSigma|^{-1/2}\exp\left[-\frac{1}{2}(\bS-\bmmu)^\top \bSigma^{-1}(\bS -\bmmu)\right]$$
where $S_i = S(\bms_i)$, $\mu_i = \mu(\bms_i)$ for some mean function $\mu(\cdot)$, and $\Sigma_{ij} =C(\bms_i,\bms_j)$ for some covariance function $C(\cdot,\cdot)$. We focus on the Mat\'ern covariance function with scaling parameter $\kappa >0$, marginal variance $\lambda^2$ and smoothness parameter $\nu$,
\begin{align}
C_\nu(\bms_i,\bms_j)  = \frac{\lambda^2}{2^{\nu -1} \gamma(\nu)}\left(\kappa||\bms_i - \bms_j||\right)^\nu K_\nu\left(\kappa ||\bms_i - \bms_j||\right) \label{eqn:matern}
\end{align}
where $||\cdot||$ denotes the Euclidean distance in $\mathbb{R}^2$ and $K_\nu$ is the modified Bessel function of the second kind and order $\nu >0$. In general, it is difficult to learn about the smoothness parameter $\nu$, and so we follow convention and fix this parameter to $\nu=1$ \citep{simpson:etal:12a,simpson:etal:12b}. The benefit of this choice is that the field has one continuous derivative while maintaining computational feasibility. We now describe how computation can be carried out for this GRF model.

In a major breakthrough an elegant connection between GRFs and GMRFs has been established
\cite{Lindgren:etal:11,simpson:etal:12a,simpson:etal:12b}. In these papers, the following stochastic partial differential equation (SPDE) is considered,
\begin{align*}
  (\kappa^2 - \Delta)^{\alpha/2}S(\bms) =\lambda W(\bms), \qquad \bms \in \mathbb{R}^2
\end{align*}
where $\Delta = (\partial^2/\partial s_1^2) + (\partial^2/\partial s_2^2)$ is the Laplacian on $\mathbb{R}^2$, $W(\bms)$ is Gaussian white noise and $\alpha = \nu + 1$. They show that the solution to the SPDE is a GRF with Mat\'ern covariance,
\begin{align*}
    S(\bms) = \int_{\mathbb{R}^2}k(\bms,\bms')~dW(\bms')
\end{align*}
where $k(\bms,\bms')=C_\nu(\bms,\bms')$.

Finally, using finite element analysis, a representation to the solution of the SPDE over a triangulation of the domain (called the mesh) is constructed by a weighted sum of basis functions,
\begin{align}
    S(\bms) \approx \tilde{S}(\bms) = \sum_{m=1}^M w_m \psi_m(\bms), \label{eqn:SPDE}
\end{align}
where $M$ is the number of mesh points in the triangulation, $\psi_m(\bms)$ is a basis function and $\bw = [w_1,\dots,w_M]^{\rT}$ is a collection of weights. The weights $\bw$ are jointly Gaussian with $\bmmu = \mathbf{0}$ and sparse $m \times m$ precision matrix, $\bQ$, depending on spatial hyperparameters $\lambda^2$ and $\kappa$; hence $\bw$ is a GMRF. The exact form for $\bQ$ is chosen so that the resulting distribution for $\tilde{S}(\bms)$ approximates the distribution of the solution to the SPDE, and thus the form will depend on the basis functions.
The basis functions are chosen to be  piecewise linear functions; that is, $\psi_m(\bms) = 1$ at the $m$-th vertex of the mesh and $\psi_m(\bms) = 0$ at all other vertices, $m=1,\dots,M$. This results in a set of pyramids, each with typically a six- or seven-sided base. % Therefore, the spatial prior consists of  functions that are weighted linear combinations of these pyramids, with the weights having a multivariate normal distribution.

%As in Chapter XX, we assume that there is a latent, continuous Gaussian random field (GRF) that varies in space, $\{S(s): s\in D \subset \mathbb{R}^2\}$ where $R$ is our study region of interest, and 
%We use the stochastic partial differential equations (SPDE) approach for the spatial modeling \citep{lindgren:etal:11}. Thus, we assume there is a latent Gaussian random field (GRF), $\{S(\bms): \bms \in D \subset \mathbb{R}^2\}$ where $D$ is the region of interest, with a Mat\'ern covariance function:
%$$\mbox{cov}(S(\bms),S(\bms') ) = \frac{\lambda^2}{2^{\upsilon-1} \Gamma(\upsilon)} (\kappa ||\bms-\bms'||)^\upsilon
%K_\upsilon(\kappa ||\bms-\bms'||),
%$$
%where $\lambda^2$ is the spatial variance, $\kappa >0$ is the scaling parameter, $K_\upsilon$ is the modified Bessel function of
%the second kind and $\upsilon > 0$ is the smoothness parameter. In general, it is difficult to learn about $\upsilon$, and so we follow convention and fix this parameter to be 1 \cite{simpson:etal:12a,simpson:etal:12b}. 

\begin{figure}
    \centering
    \includegraphics[width=0.45\linewidth]{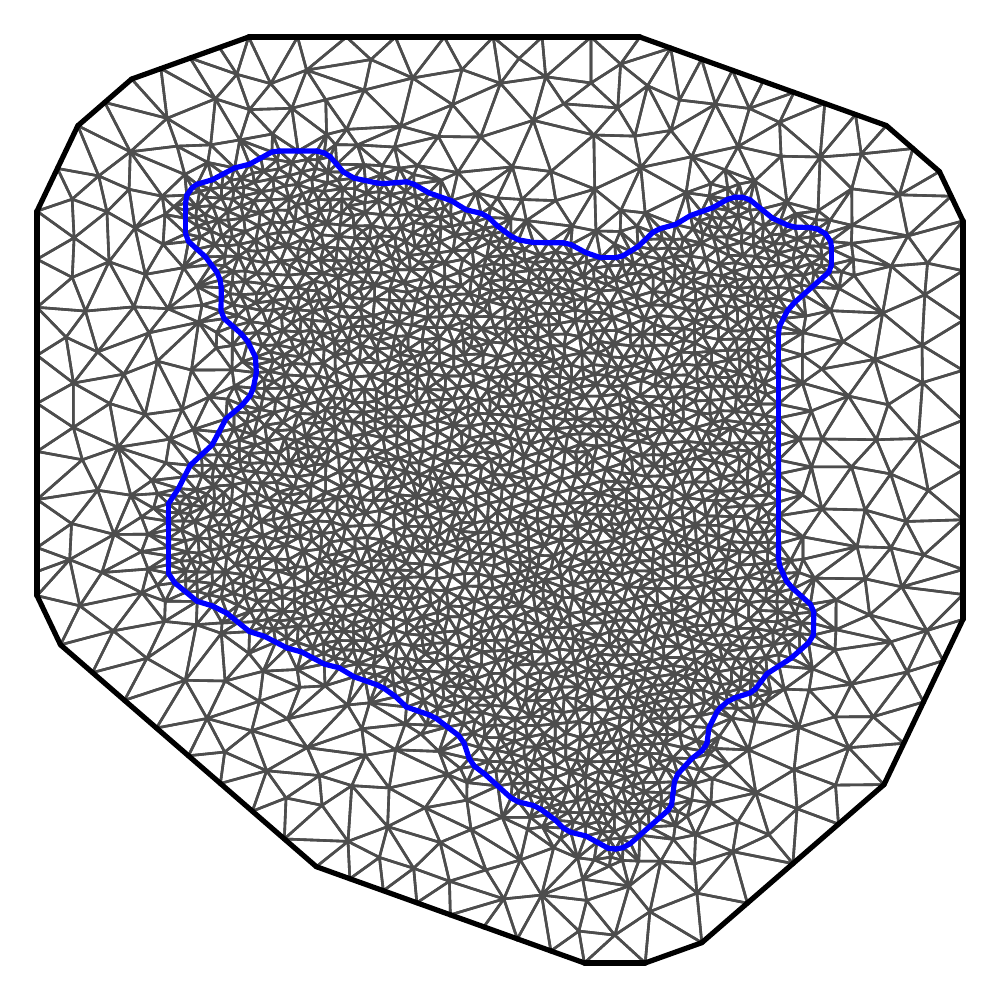}\\
    \includegraphics[width=0.475\linewidth]{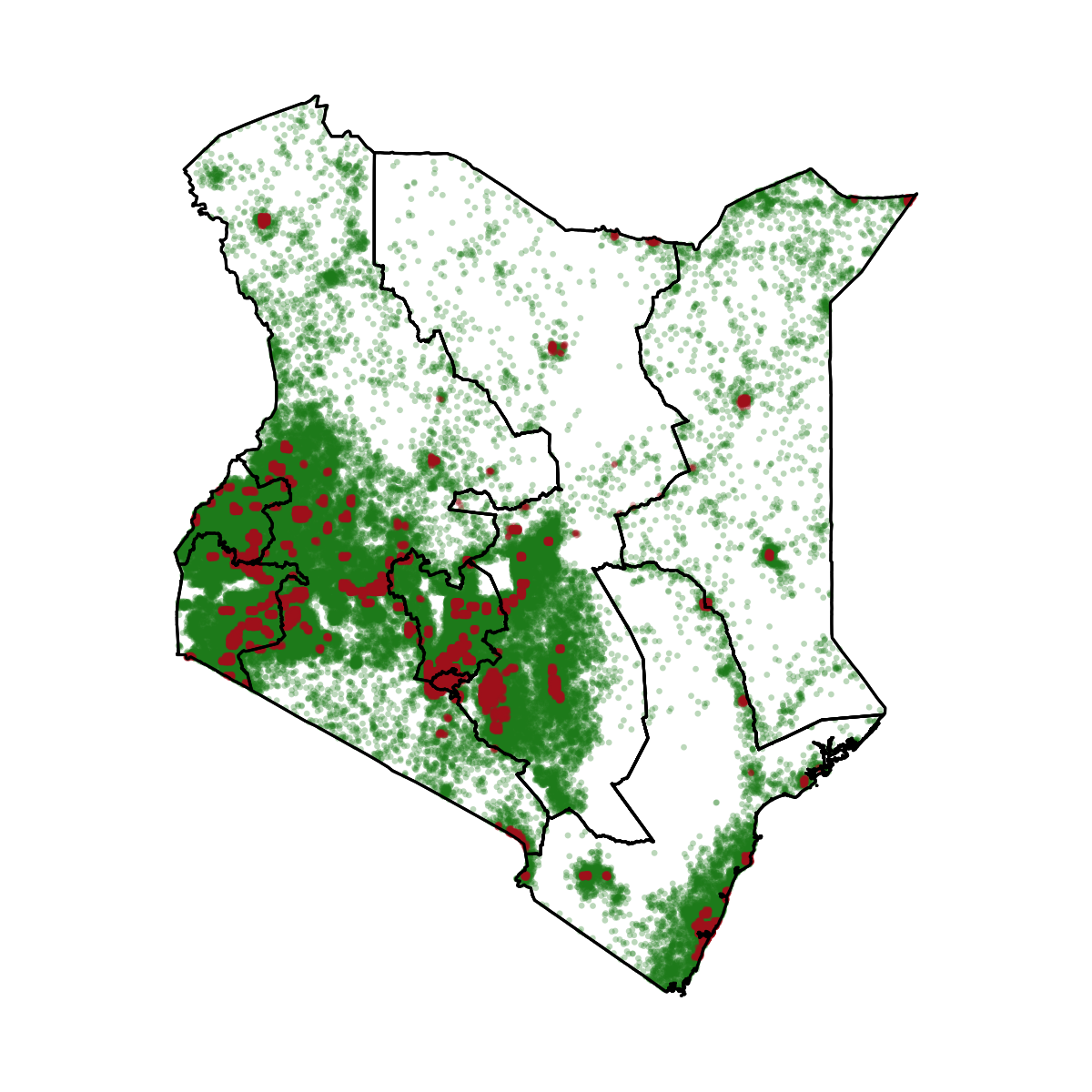}
    \includegraphics[width=0.475\linewidth]{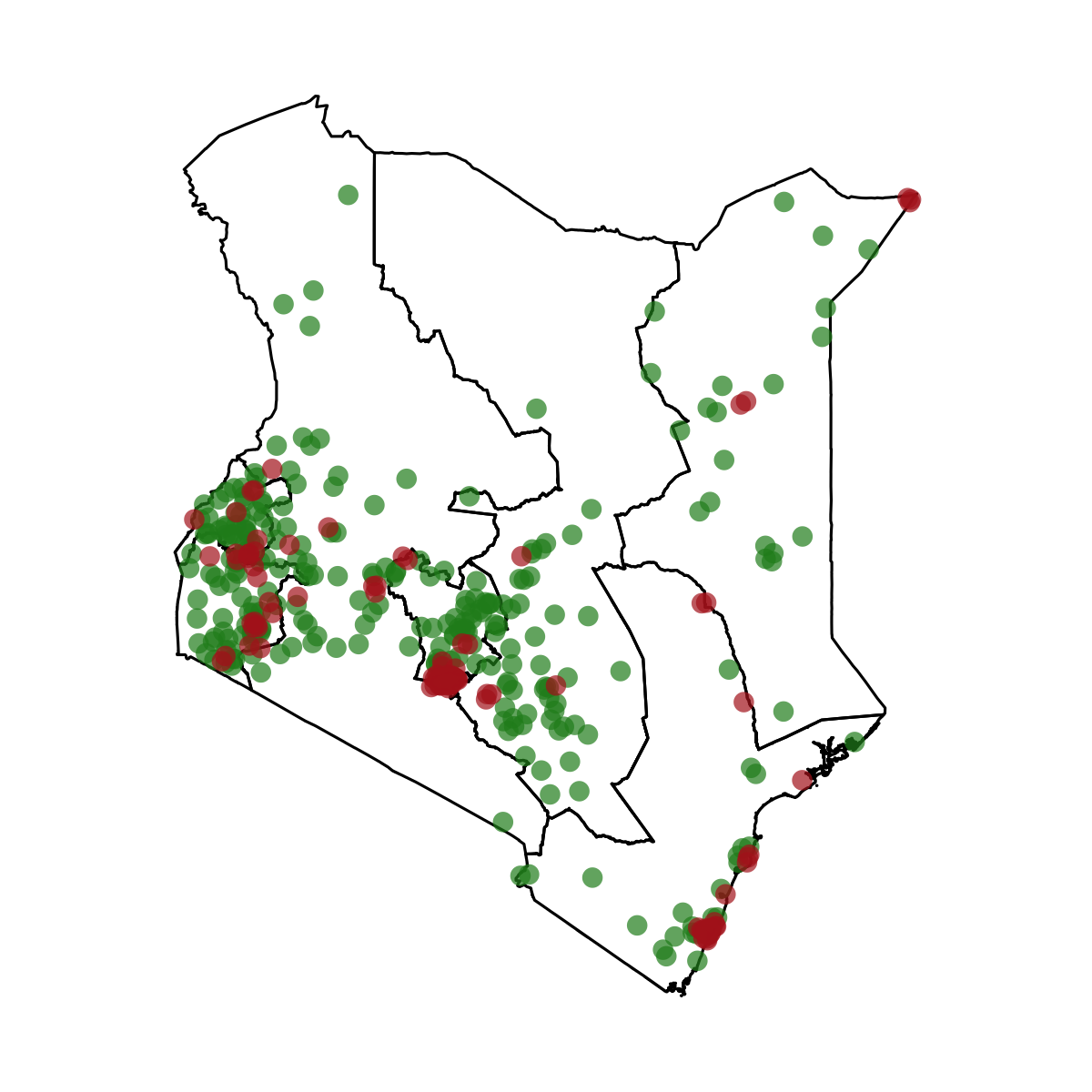}
    \caption[Mesh and locations of enumeration areas in Kenya simulation]{Top: Mesh over the geography of Kenya. Bottom left: Kenya provinces with locations of centroids. Bottom right: Kenya provinces with true locations of the 398 clusters. Red: urban. Green: rural.}
    \label{Fig:ulocs:centroids}
\end{figure}

To use the approximation, a mesh is first created. For the simulation we considered, the mesh consisted of $M=2,765$ mesh points and is shown in Figure \ref{Fig:ulocs:centroids}.
Let $\bphi=[\log \lambda, \log \kappa]$ represent the parameters of the GRF model, and $\bw$ be the vector of the weights. Let $\bbeta$ be a vector of the fixed effects, and define $\btheta = [\bbeta, \bw]$. 
We could proceed using a data augmentation (DA) algorithm, based on the factorizations:
\begin{align}
p(\btheta, \bphi | \by, \bs, \bu) \propto p(\by|\bs, \btheta, \bphi) \times p(\btheta, \bphi), \label{eq:dauloc1}\\
p(\bs | \by, \btheta, \bphi, \bu) \propto p(\by|\bs, \btheta, \bphi) \times p(\bs | \bu). \label{eq:dauloc2}
\end{align}
but this is computationally expensive, and so instead we use a hybrid scheme.

\section{INLA within MCMC}
\label{sec:comp}

For inference, we propose using an approximate Gibbs sampling strategy, known as ``INLA within MCMC'' \citep{gomez:2017b,gomez:rue:18}. The motivation for doing this is that the model described in the last section cannot be fit with INLA, given that it is a mixture distribution over the unknown locations. One could use MCMC for inference \citep{marin:etal:05}. However, such algorithms are inefficient for Gaussian processes \cite{filippone:etal:13}. The key for our implementation is to note that if the locations of the clusters are fixed, the conditional models can be fit using INLA. 

It has been recognized that some models can be fit in \texttt{R-INLA} once certain parameters in the model are fixed \cite{bivand:etal:14,bivand:etal:15} . Specifically, the authors define a grid of values for the ``problem parameters'' and use \texttt{R-INLA} to fit a conditional model. The reported marginal likelihood from \texttt{R-INLA} is then used to obtain the posterior distributions of these ``problem parameters''. Lastly, Bayesian modeling averaging \citep{hoeting:98} is used to derive the posterior distribution for the other parameters.

In the setting that we consider, it is not straightforward to derive a grid of values with high posterior probability for the ``problem parameters'' (in our case, the unknown locations). A variation on the approach of \cite{bivand:etal:14,bivand:etal:15} is to use a Metropolis-Hastings algorithm for the ``problem parameters''. This has been previously proposed \cite{gomez:2017b,gomez:rue:18}, with the marginal likelihood from fitting conditional models in \texttt{R-INLA} being used to determine the acceptance probability for the Metropolis-Hastings step. They note that the marginal likelihood reported by \texttt{R-INLA} is an estimate, and the limiting distribution is not exactly the desired stationary distribution. For their purposes, they argue and show that the difference is not significant.

We take this approach one step further and propose a new algorithm using \texttt{R-INLA} to fit (\ref{eq:dauloc1}) and generate a sample for $[\btheta, \bphi]$ based on the posterior (INLA) approximation. This sample is then used to generate a sample for $\bms$. This avoids needing a proposal distribution, as the posterior conditional distribution is available exactly. 
Therefore, the ``INLA within MCMC'' algorithm can be summarized as:
\begin{enumerate} %[topsep=0pt,itemsep=-1ex,partopsep=1ex,parsep=1ex]
    \item Initialize $\btheta^{(0)} = [\bbeta^{(0)},\bw^{(0)}]$, where $\bw$ represents the values of the spatial field on the mesh.
    \item Iterate:
    \begin{enumerate}
        \item Sample $\bms_{ijk}^{(t+1)}$ using Gibbs sampling,
        \begin{align}
        p(\bms_{ijk} = \bmE_{ije}|y_{ijk},\bmu_{ijk},\btheta^{(t)}) & \propto p(\bms_{ijk} = \bmE_{ije} | \bmu_{ijk}) \times p(y_{ijk} | \bms_{ijk}=\bmE_{ije}, \btheta^{(t)}) \label{eqn:ulocs:step1}
        \end{align}
        where the first term on the right corresponds to (\ref{eq:maskprior}) for the masking scenario and (\ref{eqn:jitterprior}) for the displacement scenario. The second term on the right corresponds to the complete data likelihood.
        \item Use INLA to obtain the approximate conditional posterior, denoted $\tilde{p}(\btheta, \bphi|\by,\bs^{(t+1)})$. Sample $\btheta^{(t+1)}, \bphi^{(t+1)}$ from the approximate posterior.
    \end{enumerate}
\end{enumerate}

We note that the implementation of \texttt{R-INLA} means that the hyperparameters, $\bmphi$, are defined on a grid meaning that when a joint sample is drawn from the approximation in step (b) the hyperparameters can only fall on the grid. By default, a new grid for the hyperparameters is constructed during each iteration (since the cluster locations changes). This grid could be fixed ahead of time if this is desired; however, we allow the grid to change from one iteration to the next in our examples.
How accurate the results are relies on the accuracy of INLA and on the joint posterior sampling algorithm used in \texttt{R-INLA}.% [UNCLEAR ABOUT THIS].  
The latter is based on a mixture of multivariate normal distributions, with the mixing being over the grid of hyperparameters.

\section{Simulation Setup}
\label{sec:sim}

We investigate the impact of masking and displacement of cluster centroids using the geography of Kenya. A masterframe of all sampling locations approximately representing the true masterframe from the 2009 Kenya census was created based on population density retrieved from \cite{worldpop:population:16}. This was done by first dividing the gridded population density into two zones: urban and rural within each county. To identify these zones, thresholding was used so that the proportion exceeding the threshold amount matched the proportion urban in the 2014 Kenya DHS \citep{KenyaDHS:15}. The 1km by 1km grids that exceeded the threshold were labeled as urban and otherwise labeled rural. The masterframe of all sampling locations was then created by randomly drawing coordinates proportional to population density within each strata (urban/rural crossed with county) to obtain 95,310 enumeration areas; see Figure \ref{Fig:ulocs:centroids} for locations and Table \ref{tab:ulocs:masknumlocs} for counts of clusters by each of the eight provinces and the urban/rural strata. Finally, 398 clusters (right panel in Figure \ref{Fig:ulocs:centroids}) were then randomly sampled (uniformly, not proportional to size), stratified by province and urban/rural. The number of clusters within each sampling strata were chosen to match the 2008 Kenya DHS.

\begin{table}[tbp]
    \centering
    \begin{tabular}{l|rr}
        & Rural & Urban \\ \hline
        Central & 7,816 & 4,192 \\
        Coast & 4,268 & 3,569 \\
        Eastern & 12,396 & 3,234 \\
        Nairobi & 0 & 10,394 \\
        North Eastern & 2,230 & 433 \\
        Nyanza & 9,787 & 3,041 \\
        Rift Valley & 19,097 & 6,051 \\
        Western & 7,383 & 1,419
    \end{tabular}
    \caption{Number of potential clusters in each administrative area and strata.}
    \label{tab:ulocs:masknumlocs}
\end{table}

\subsection{Model}

\begin{figure}
    \centering
    \includegraphics[width=0.45\linewidth]{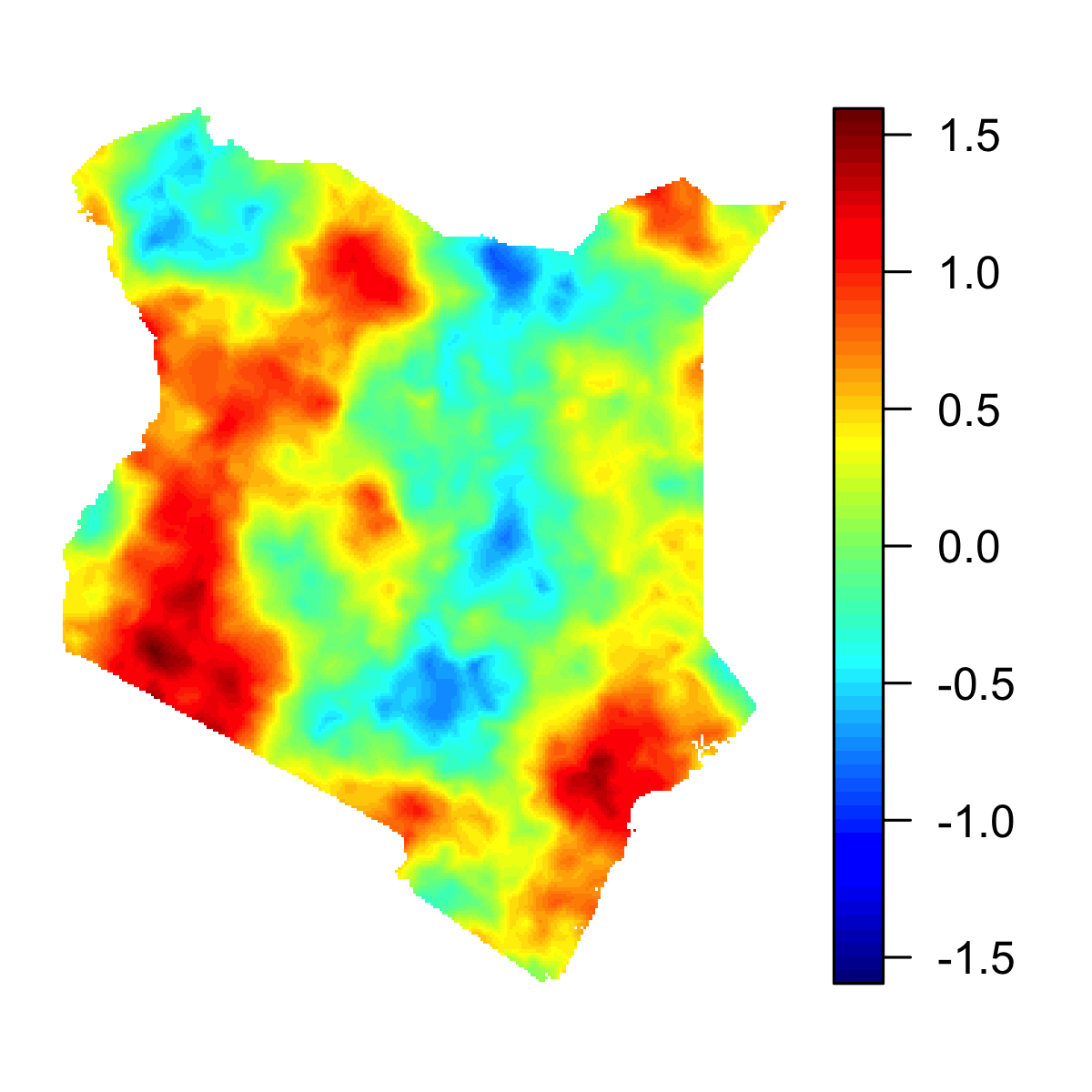}
    \includegraphics[width=0.45\linewidth]{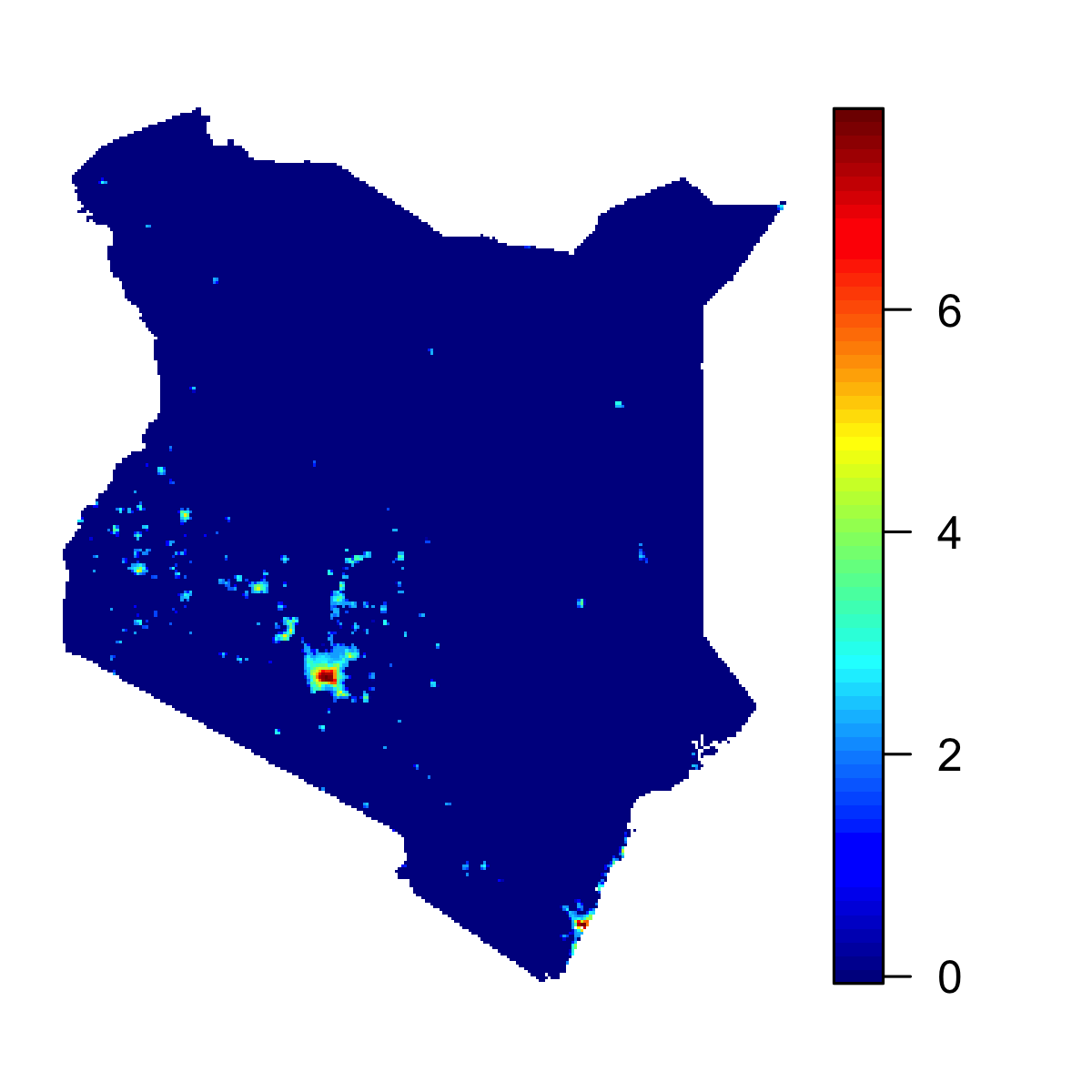}
    \caption[Latent spatial surface and covariate surface used in simulation]{Left: Spatial surface $\tilde{S}(\cdot)$ used in the simulation. Right: square root of nighttime lights surface.}
    \label{fig:ulocs:surf_and_lights}
\end{figure}

We imagine a binary response and generate 
data from the model,
\begin{align*}
    Y_{ijk}| p(\bms_{ijk}) & \sim \text{Binomial}(25,  p(\bms_{ijk}))\\
    \text{logit}(p(\bms_{ijk})) & = \beta_0 + \beta_1 z_{ijk} + \tilde{S}(\bms_{ijk})
\end{align*}
where $\tilde{S}(\cdot)$ is the SPDE approximation to the Gaussian process spatial random effect surface. In our simulation, we used the square-root of nighttime lights \nocite{nightimelights} 
(NOAA nighttime lights series) as the spatial covariate, $\bz$. The spatial surface $\tilde{S}(\cdot)$ and nighttime lights surface are plotted in Figure \ref{fig:ulocs:surf_and_lights}.

\subsection{Scenarios}

\begin{figure}
    \centering
    \includegraphics[width=0.4\linewidth]{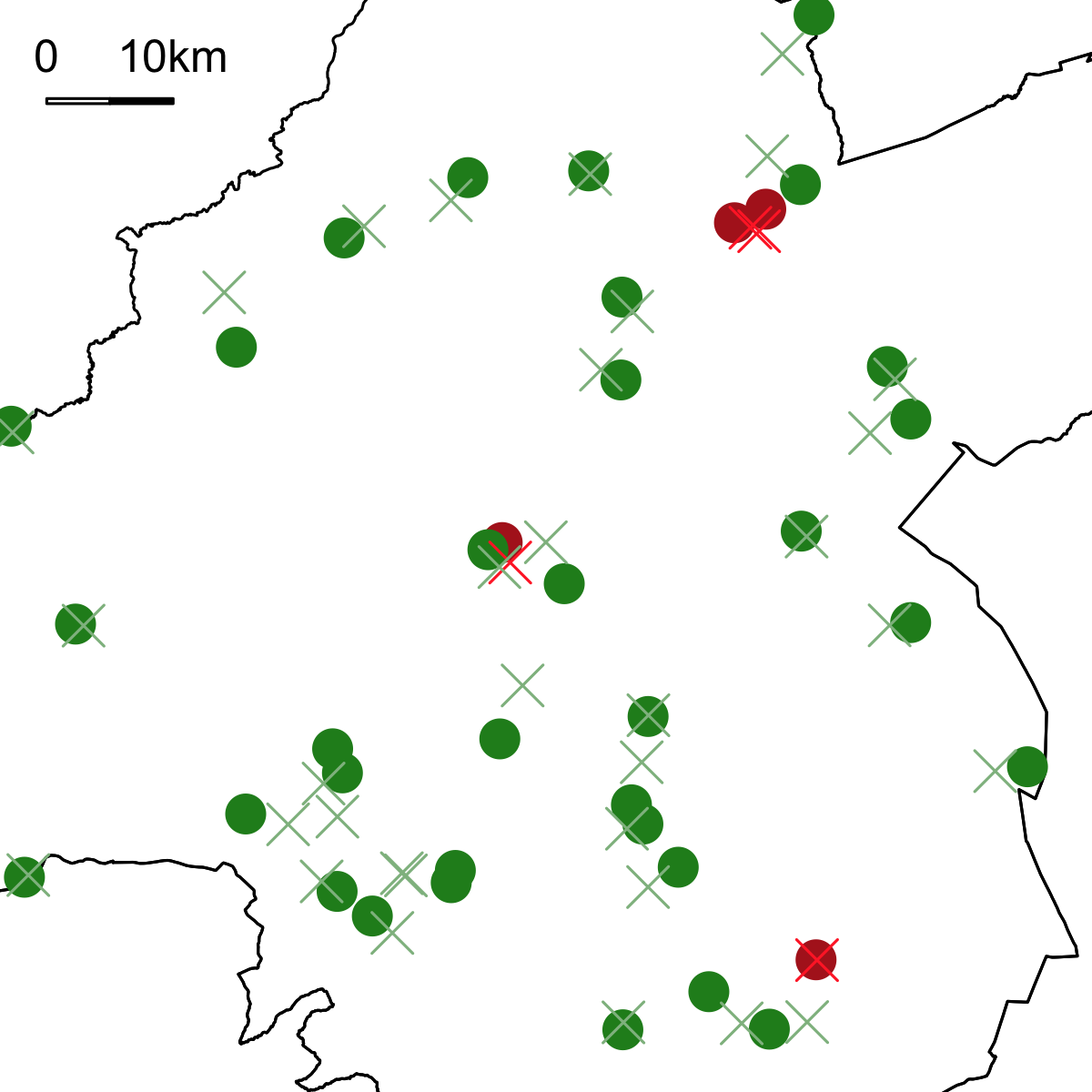}
    \hspace{5mm}
    \includegraphics[width=0.475\linewidth]{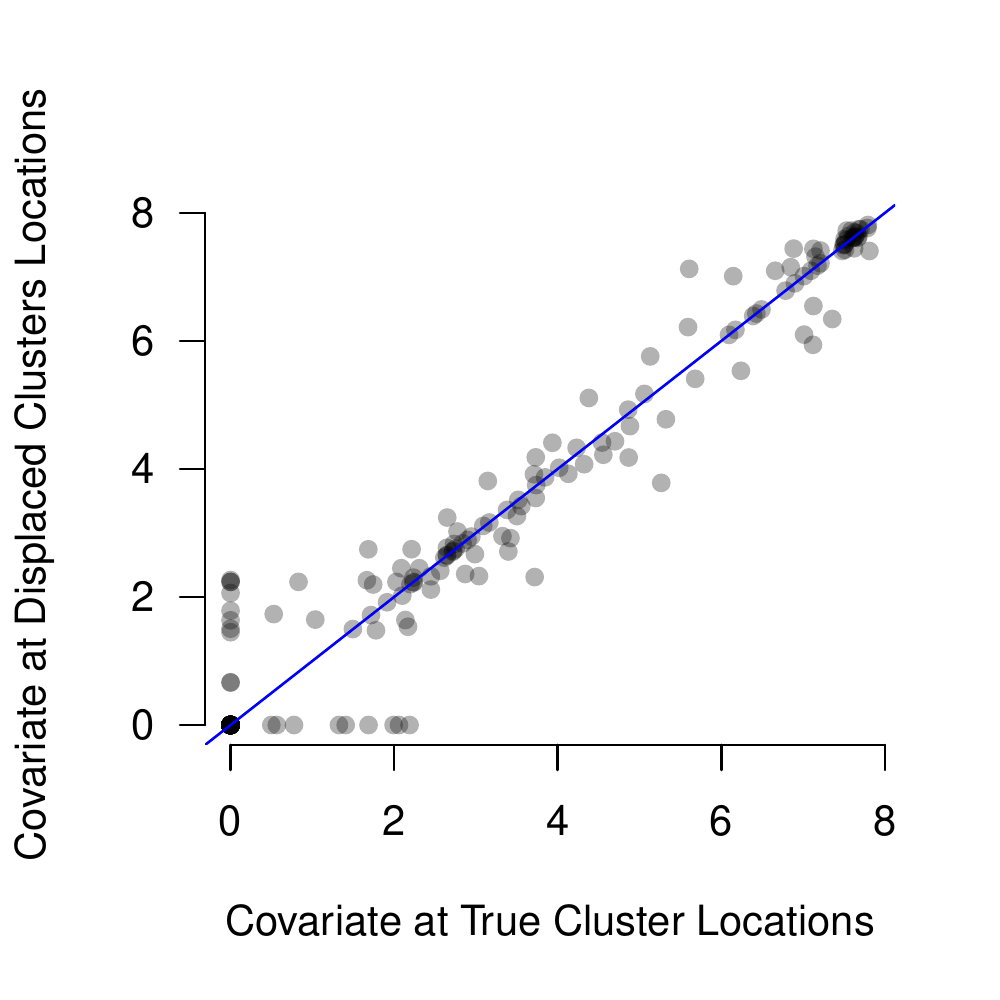}
    \caption[True and jittered locations and covariate values in simulation]{Left: true and jittered locations in simulation, zoomed in on the Western province. Solid points: true locations of clusters. {$\times$}: displaced locations. Red: urban clusters. Green: rural clusters. Right: value of covariate at true and jittered locations.}
    \label{fig:ulocs:jitteredvstrue}
\end{figure}

\begin{figure}
    \centering
    \includegraphics[width=0.5\linewidth]{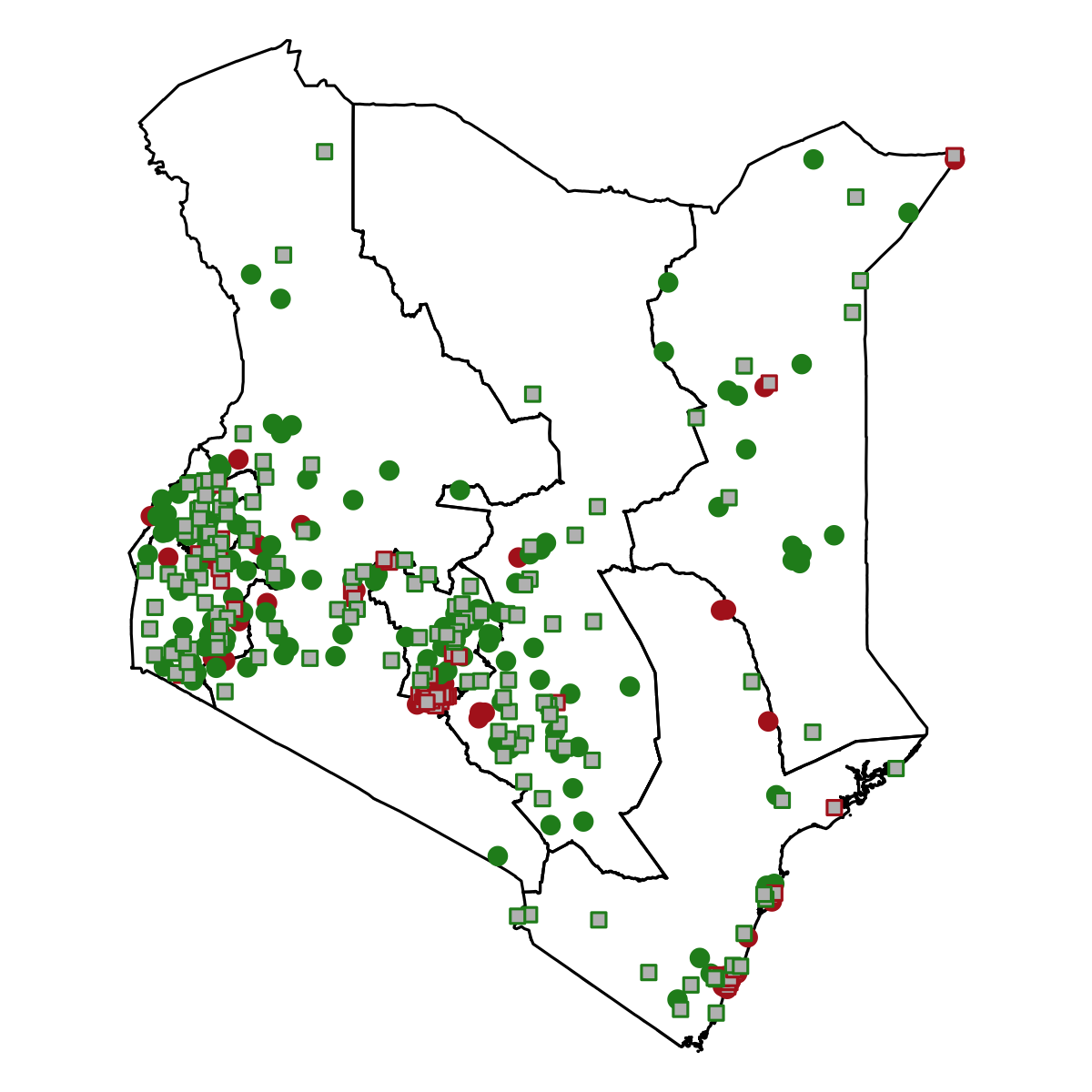}
    \caption[Cluster location availability for masked scenarios]{Solid points: GPS locations of clusters known. Grey squares: only admin area of clusters known. i.e.,~masked data. Red: urban clusters. Green: rural clusters.}
    \label{fig:ulocs:masked}
\end{figure}

We will consider several simulation scenarios where centroid locations are jittered or masked, described in Table \ref{tab:uloc-sim-scen}. Row 1a corresponds to fitting the ``gold standard'' model, which is if all cluster centroids are available exactly. Row 2a corresponds to fitting the model using the jittered locations of the centroids. Figure \ref{fig:ulocs:jitteredvstrue} shows the true and jittered locations for clusters in the Western province. Row 3a corresponds to using our proposed approach to accommodate the jittered nature of the locations. Rows 4a--6a refer to a masking scenario where we will mask 50\% of the centroids (so that only the strata and administrative area are known); Figure \ref{fig:ulocs:masked} shows the location of the clusters where the true centroids are known and only the administrative area and strata are known. Row 4a corresponds to fitting the model only to the data where the cluster information is known exactly. Row 5a corresponds to also incorporating the data from the masked cluster locations. Here, we use the centroid of all potential cluster locations.
Row 6a corresponds to using our proposed approach. We repeat these scenarios when a covariate is included (rows 1b--6b). In each case, we investigate the effect on surface reconstruction and covariate associations.

\begin{table}[tbp]
    \centering
    \begin{tabular}{cc|ccc}
    &  & Centroid Locations & Spatial Covariate\\
    \hline
    1a & INLA &  100\% exact & \\
    2a & INLA naive & 100\% jittered & \\
    3a & INLA within MCMC & 100\% jittered & \\
    4a & INLA & 50\% exact & \\
    5a & INLA & 50\% exact, 50\% at centroids & \\
    6a & INLA within MCMC & 50\% exact, 50\% masked & \\
    1b & INLA &  100\% exact & \checkmark \\
    2b & INLA naive & 100\% jittered & \checkmark \\
    3b & INLA within MCMC & 100\% jittered & \checkmark \\
    4b & INLA & 50\% exact & \checkmark \\
    5b & INLA & 50\% exact, 50\% at centroids & \checkmark \\
    6b & INLA within MCMC & 50\% exact, 50\% masked & \checkmark
    \end{tabular}
    \caption{Simulation scenarios considered.}
    \label{tab:uloc-sim-scen}
\end{table}

\subsection{Computation}
\label{sec:ulocs:sim:comp}

For the ``INLA within MCMC'' algorithm, (\ref{eqn:ulocs:step1}) is as follows:
\begin{align*}
            p(\bms_{ijk}^{(t+1)} = \bmE_{ije}|y_{ijk},\bmu_{ijk},\btheta^{(t)}) & \propto p(\bms_{ijk}^{(t+1)} = \bmE_{ije} | \bmu_{ijk}) \times \\
            & \qquad \left\{\text{expit}\left(\beta_0^{(t)} +  \beta^{(t)}_1 z(\bmE_{ije}) + \tilde{S}(\bmE_{ije})^{(t)}\right)\right\}^{y_{ijk}} \times \\
            & \qquad \left\{1-\text{expit}\left(\beta^{(t)}_0 +   \beta^{(t)}_1 z(\bmE_{ije})  + \tilde{S}(\bmE_{ije})^{(t)}\right)\right\}^{25 - y_{ijk}}
        \end{align*}
where $\btheta = (\bbeta,\bw)$, $\bms_{ijk}$ is the true location of cluster $k$ in strata $j$ and administrative area $i$, $\bmu_{ijk}$ is the available location information for the cluster, and $\bmE_{ije}$ is a potential cluster location in strata $j$ and area $i$, $e=1,\dots,m_{ij}$.
The number of potential locations, $m_{ij}$, ranged from 1 to 1,015 for the jittering scenario (median 142) and 433 to 19,097 for the masking scenario. To obtain the normalization factors (\ref{eq:ulocs:norm}), for each possible enumeration area, we simulate 1,000 jitterings of the point following the DHS jittering algorithm and determine the proportion of realizations that fall within the administrative area.
The priors for the fixed effects (intercept and covariate association) were $N(0,100)$. The hyperprior for $\bphi= [\log \lambda,~ \log \kappa]^\top$ is chosen to be fairly vague. Here, the prior mean for $\phi_1$ corresponds to a marginal variance $\lambda^2$ of 1. 
The prior mean for $\phi_2$ corresponds to a practical range of roughly 20\% of the domain size. 
%This leads to an effective range of 255 km and variance of the spatial process was set to 0.25.  
Code to fit the models can be found at \url{https://github.com/wilsonka/Incomplete-Geography}
%The priors for these spatial hyperparameters were analogous to the ones given in Section \ref{sec:pointless:comp}. 

\section{Simulation Results}
\label{sec:simres}

\begin{figure}[tbp]
    \centering
    \includegraphics[width=0.8\linewidth]{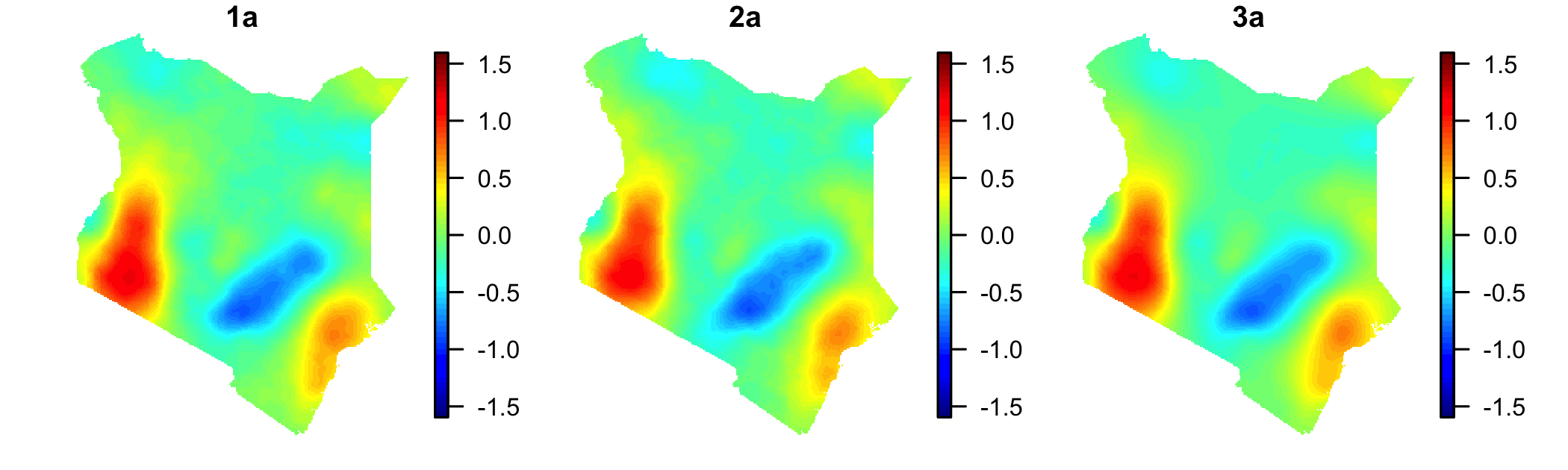}\\
    \includegraphics[width=0.8\linewidth]{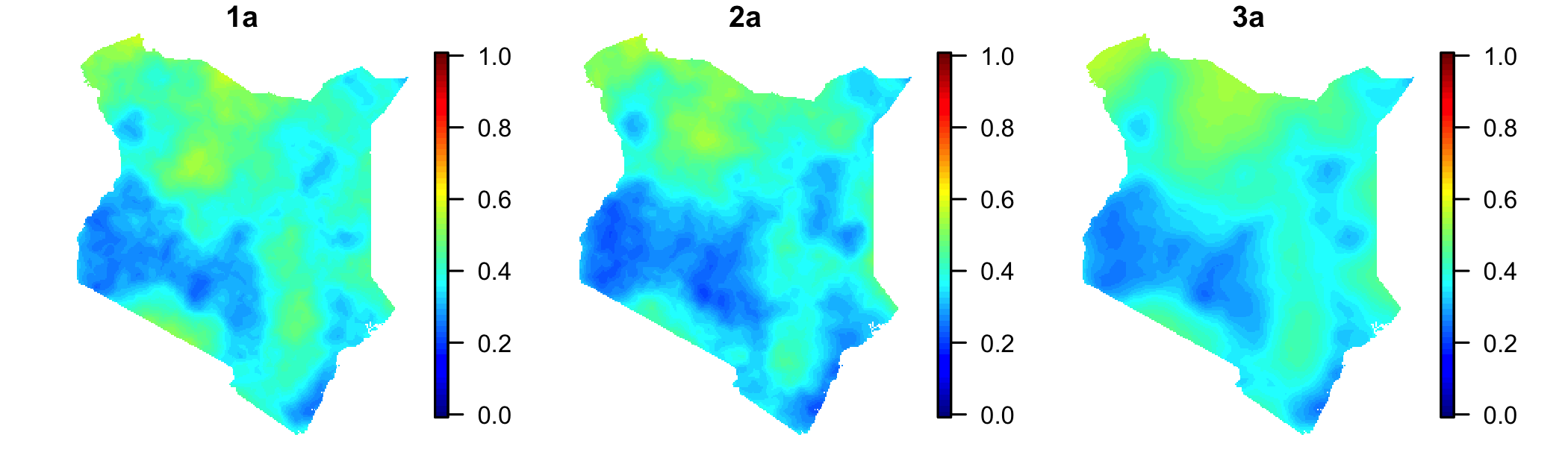}
    \caption[Posterior medians and standard deviations of latent spatial surface for the jittering scenario without covariate]{Top row: posterior medians of latent spatial surface. Bottom row: posterior standard deviations of latent spatial surface for the jittering scenario without a spatial covariate. The left column is the ideal scenario with exact GPS available (see Table \ref{tab:uloc-sim-scen} for full  description of the different scenarios).}
    \label{fig:ulocs:surfacejitter}
\end{figure}

\begin{figure}[tbp]
    \centering
    \includegraphics[width=0.8\linewidth]{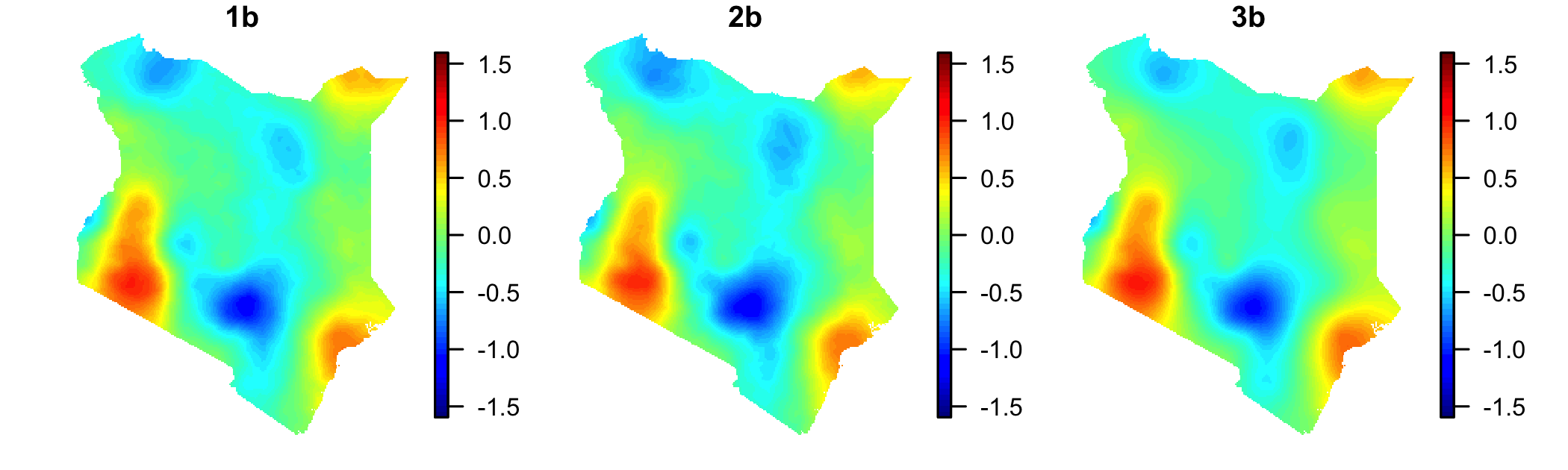}\\
    \includegraphics[width=0.8\linewidth]{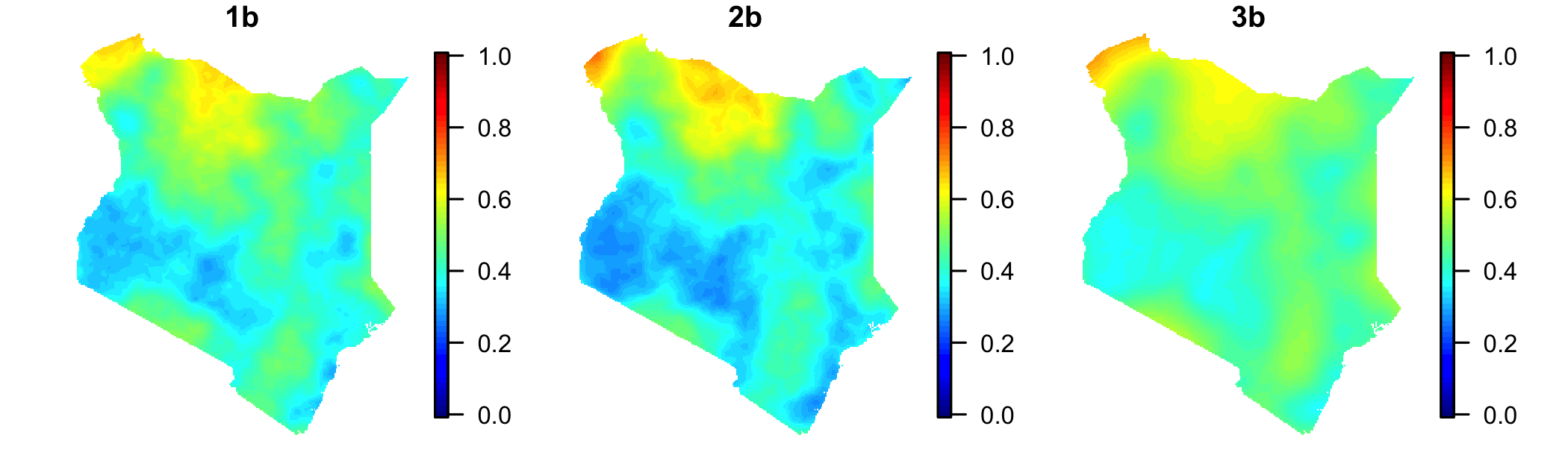}
    \caption[Posterior medians and standard deviations of latent spatial surface for the jittering scenario with covariate]{Top row: posterior medians of latent spatial surface. Bottom row: posterior standard deviations of latent spatial surface for the jittering scenario with a spatial covariate. The left column is the ideal scenario with exact GPS available (see Table \ref{tab:uloc-sim-scen} for full description of the different scenarios).}
    \label{fig:ulocs:surfacejittercov}
\end{figure}

\begin{figure}[tbp]
    \centering
    \includegraphics[width=0.8\linewidth]{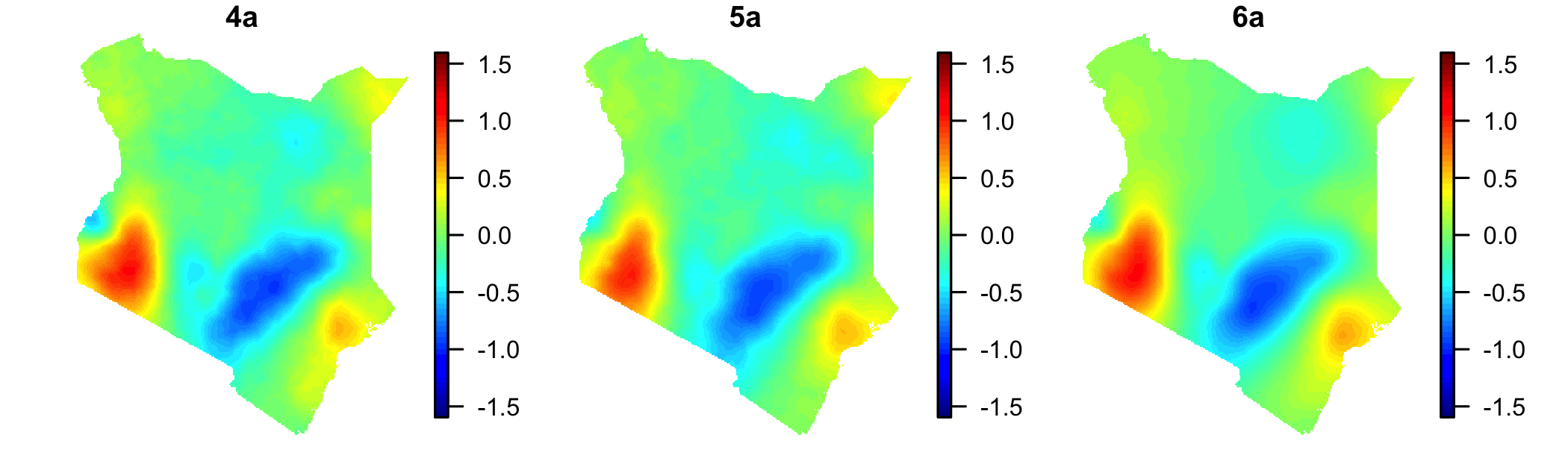}\\
    \includegraphics[width=0.8\linewidth]{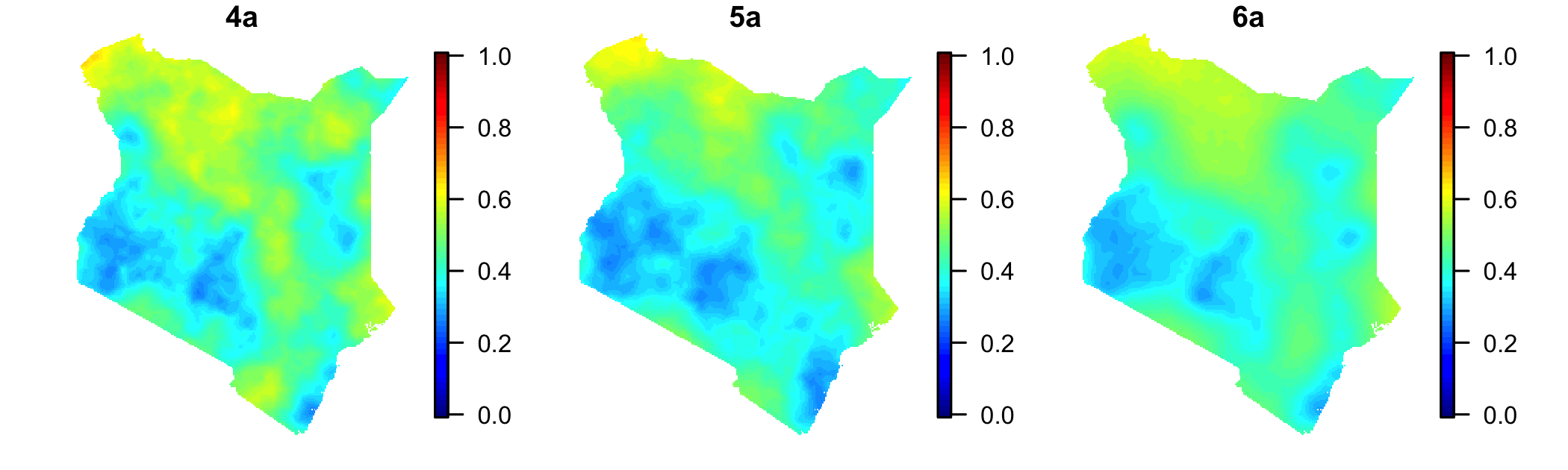}
    \caption[Posterior medians and standard deviations of latent spatial surface for the masking scenario without covariate]{Top row: posterior medians of latent spatial surface. Bottom row: posterior standard deviations of latent spatial surface for the masking scenario without spatial covariate.}
    \label{fig:ulocs:surfacemasknocov}
\end{figure}

\begin{figure}[tbp]
    \centering
    \includegraphics[width=0.8\linewidth]{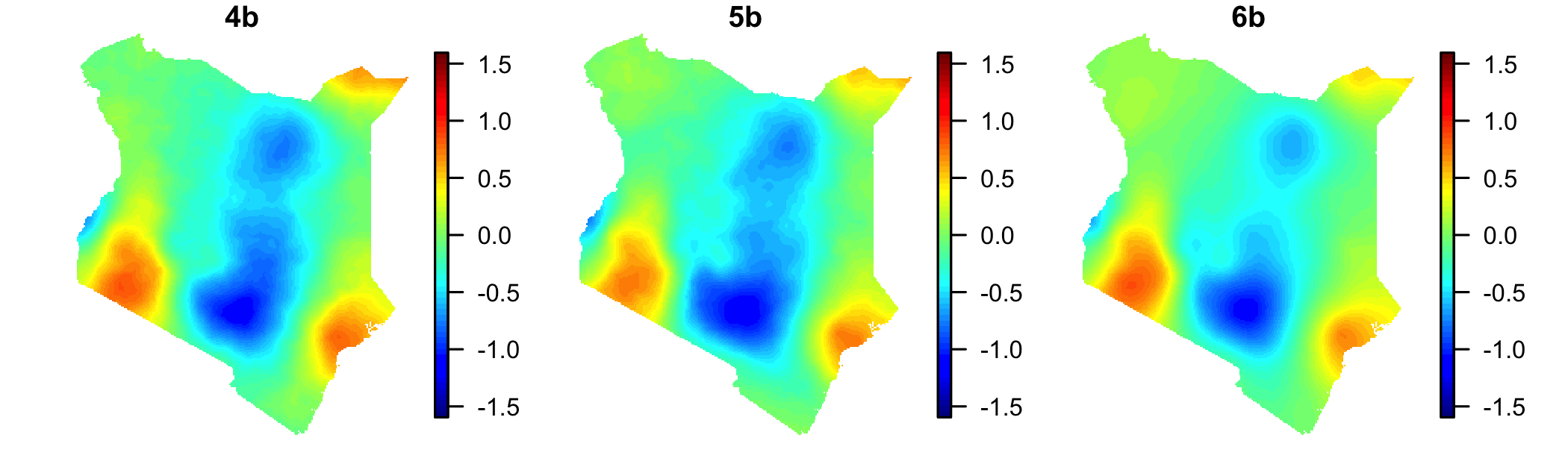}\\
    \includegraphics[width=0.8\linewidth]{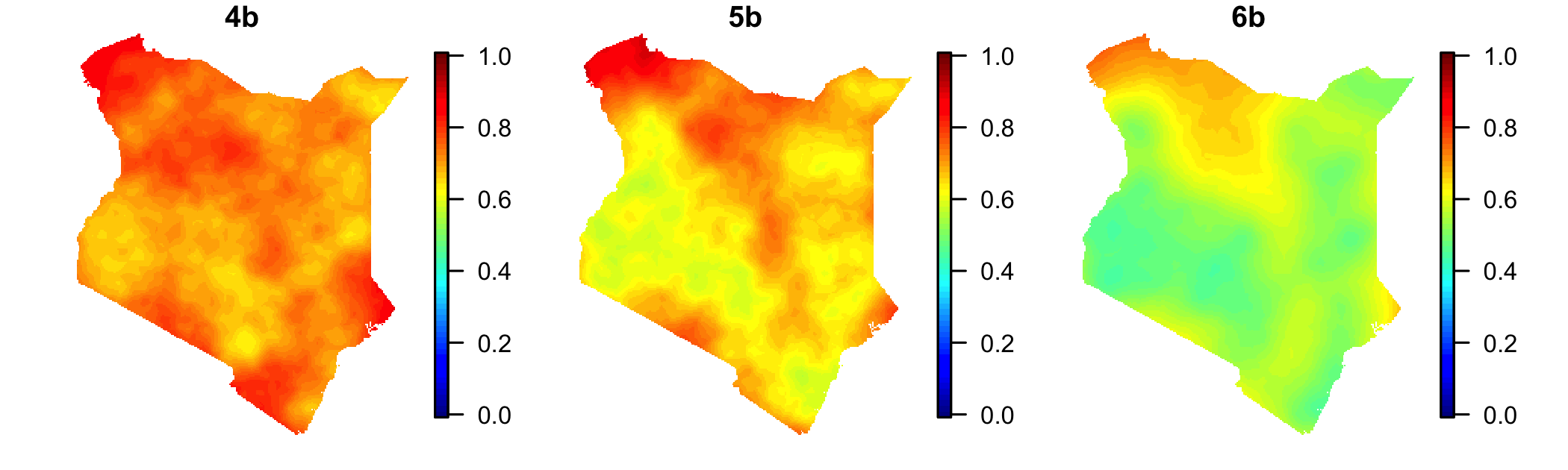}
    \caption[Posterior medians and standard deviations of latent spatial surface for the masking scenario without covariate]{Top row: posterior medians of latent spatial surface. Bottom row: posterior standard deviations of latent spatial surface for the masking scenario with spatial covariate.}
    \label{fig:ulocs:surfacemaskcov}
\end{figure}

To assess convergence, trace plots were examined and we calculated the $\hat{R}$ statistic \citep{gelman:rubin:92} and these were all less than 1.05, which suggests convergence for all scenarios and approaches.
Posterior medians and 95\% credible intervals (CIs) for the fixed effects and spatial hyperparameters are presented in Table \ref{tab:ulocs:results}. Figures \ref{fig:ulocs:surfacejitter}--\ref{fig:ulocs:surfacemaskcov} shows the posterior medians latent surface $\tilde{S}(\cdot)$  and posterior standard deviation for the jittering and masking scenarios, respectively. First, we consider jittering, where the ``best case'' scenarios are 1a and 1b, where the true cluster locations are available. In general, using the jittered coordinates does not significantly impact the results, except for differences in the uncertainty in the spatial surface (bottom rows of Figures \ref{fig:ulocs:surfacejitter} and \ref{fig:ulocs:surfacejittercov}). Additionally, the differences are larger when a spatial covariate is involved in our simulation (1b and 2b). When using the DA approach for jittered data (3a and 3b), we also see some minor differences, and some ``recovery'' of the best case scenario. 

\begin{table}[tbp]
    \centering
    \footnotesize
    \begin{tabular}{c|cccc}
    Model & $\beta_0$ & $\beta_1$ & $\phi_1$ & $\phi_2$\\
    \hline
    Truth & -1.5 & 0.15 & 3.93 & -4.5 \\
    \hline
    1a & -1.24 (-1.75, -0.85) & - & 3.97 (3.60, 4.36) & -4.55 (-5.09, -4.05) \\
    2a & -1.22 (-1.64, -0.84) & - & 3.97 (3.60, 4.36) & -4.55 (-5.09, -4.05) \\
    3a & -1.23 (-1.69, -0.81) & - & 3.98 (3.61, 4.35) & -4.58 (-5.22, -4.10) \\
    4a & -1.09 (-1.43, -0.50) & - & 3.81 (3.33, 4.31) & -4.45 (-5.08, -3.86) \\
    5a & -1.07 (-1.58, -0.55) & - & 4.03 (3.57, 4.51) & -4.61 (-5.25, -4.00) \\
    6a & -1.13 (-1.59, -0.63) & - & 3.94 (3.44, 4.40) & -4.56 (-5.35, -4.00) \\
    1b & -1.10 (-1.67, -0.48) & 0.16 (0.12, 0.19) & 4.00 (3.63, 4.40) & -4.71 (-5.36, -4.13) \\
    2b & -1.08 (-1.58, -0.59) & 0.14 (0.11, 0.18) & 3.98 (3.61, 4.39) & -4.69 (-5.34, -4.11) \\
    3b & -1.14 (-1.81, -0.48) & 0.16 (0.12, 0.19) & 3.99 (3.60, 4.37) & -4.71 (-5.52, -4.16) \\
    4b & -1.06 (-1.99, 0.05) & 0.18 (0.13, 0.23) & 4.02 (3.56, 4.52) & -4.82 (-5.59, -4.13)\\
    5b & -1.00 (-1.89, 0.27) & 0.19 (0.15, 0.23) & 4.05 (3.59, 4.54) & -4.85 (-5.62, -4.15)\\
    6b & -1.05 (-1.81, -0.27) & 0.18 (0.14, 0.23) & 4.05 (3.57, 4.51) & -4.79 (-5.74, -4.17)
    \end{tabular}
    \caption{Posterior medians (95\% CIs) for parameters in the simulation scenarios considered (see Table \ref{tab:uloc-sim-scen} for description of the different scenarios).}
    \label{tab:ulocs:results}
\end{table}

Next, we consider the masking scenario, where the ``best case'' scenarios are again 1a and 1b, where true cluster locations are available for all clusters. Across the board, posteriors tend to be wider when we consider cases where only 50\% of the clusters with GPS coordinates (4a and 4b). The approach that uses the centroid for the masked data (5a and 5b) gives slightly different results. Noticeably, these results tend to be worse when a spatial covariate is involved (5b). In this scenario, we had taken the location for the masked data to be the centroid location of the potential locations and  used the value of the spatial covariate at that centroid location (rather than averaging the covariate from the potential locations). The DA approach (6a and 6b), where the other 50\% of the clusters with only the admin area known are also included, show similar results. We find a more noticeable narrowing of the 95\% CIs in the scenario involving the covariate (6b).

\begin{sidewaysfigure}
\centering
\includegraphics[width=\linewidth]{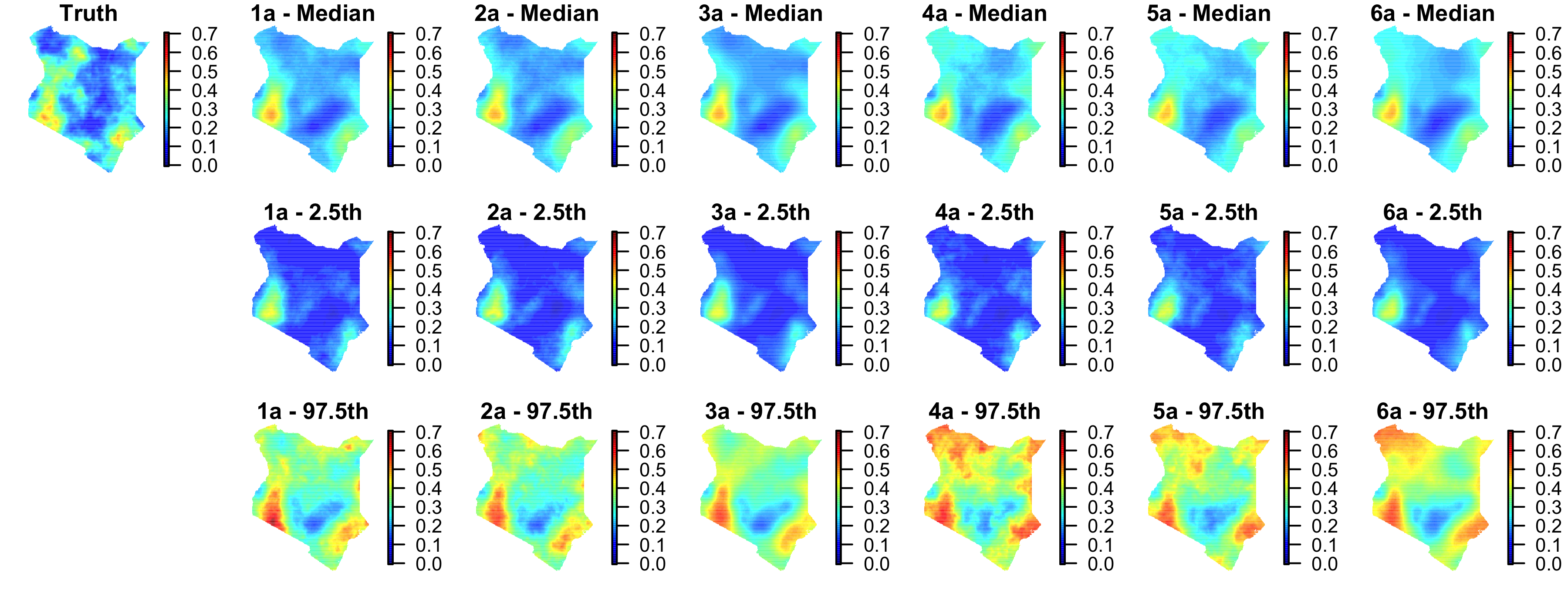}
\caption[Predicted probability surface for simulation without spatial covariate.]{Predicted probability surface for simulation without spatial covariate. Top row: posterior median. Middle row: 2.5th percentile. Bottom row: 97.5th percentile.}
\label{fig:ulocs:pnocov}
\end{sidewaysfigure}

\begin{sidewaysfigure}
\centering
\includegraphics[width=\linewidth]{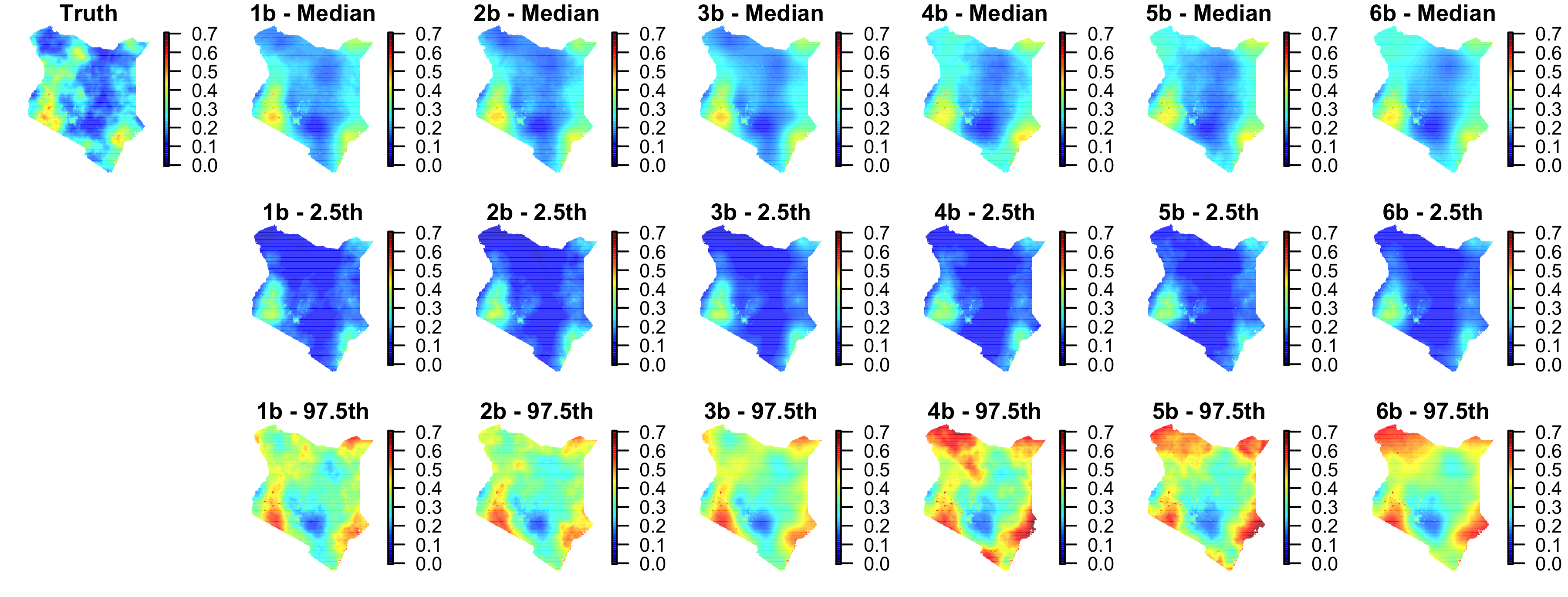}
\caption[Predicted probability surface for simulation with spatial covariate.]{Predicted probability surface for simulation with spatial covariate. Top row: posterior median. Middle row: 2.5th percentile. Bottom row: 97.5th percentile.}
\label{fig:ulocs:pcov}
\end{sidewaysfigure}

\begin{figure}
    \centering
    \includegraphics[width=0.8\linewidth]{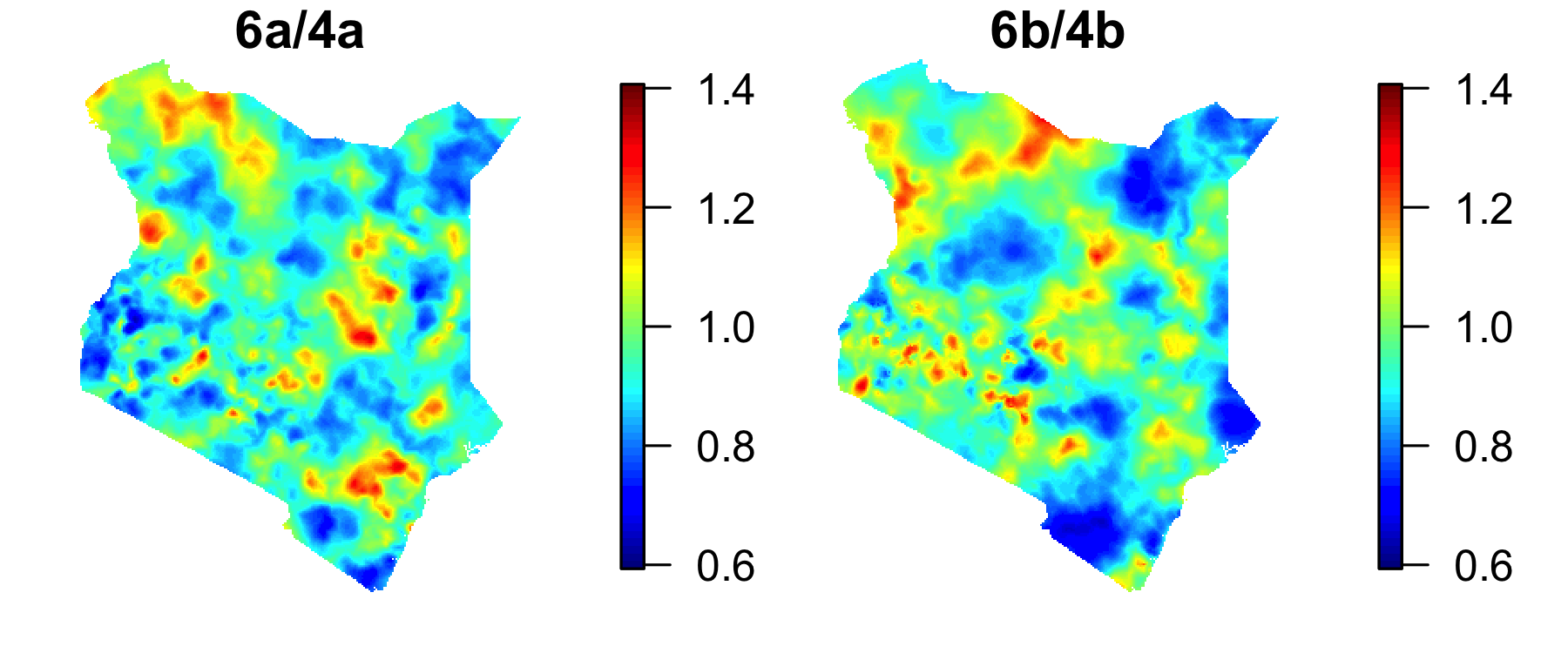}
    \caption[Ratio of posterior standard deviation of logit($p$) in DA approach to 50\% only approach for masking scenario.]{Ratio of posterior standard deviation of logit($p$) in DA approach to 50\% only approach for masking scenario. Values less than 1 indicates that the posterior standard deviation is lower when the data with masked location information is incorporated over not including it.}
    \label{fig:ulocs:sdcomp}
\end{figure}

The predicted probability surfaces are in Figures \ref{fig:ulocs:pnocov} and \ref{fig:ulocs:pcov}. Plotted are the posterior medians and 95\% CIs. The posterior medians tend to be similar within the jittering scenarios and within the masking scenarios. There is some overall reduction in uncertainty when using DA for the masking scenario, though this varies significantly spatially (Figure \ref{fig:ulocs:sdcomp}).

Additionally, we consider the mean squared error (MSE) of the predicted latent surface $S(s)$ and the probability surface (on the logit scale) over Kenya, 
$$\text{MSE}^{(M)} = \frac{1}{G}\sum_{g=1}^G \left\{E(Y_g^{(M)} - y_g)\right\}^2 + \frac{1}{G}\sum_{g=1}^G \text{Var}(Y_g^{(M)})$$
with $g$ indexing points, $y_g$ being the true value of the surface at location $s_g$, and $Y_g^{(M)}$ being the estimate for the surface for model $M$ at location $g$. We consider 2 different resolutions. In the first, predictions are made on a 1km $\times$ 1km grid, i.e., the grid points are $1$km apart. In the second, the predicted probability surface on the 1km $\times$ 1km grid is aggregated up to obtain predictions on a 5km $\times$ 5km grid. The values for the 1km $\times$ 1km grid, including average squared bias are in Table \ref{tab:ulocs:mse} (results were similar for 5km $\times$ 5km) and we can see that using the reported locations (the naive approach) tends to result in more bias as compared to using the correct approach, though this does not always hold. There also seems to be little to no benefit in using the DA approach in this setting. However, when we consider the masking scenario, we find a benefit in using DA over including the masked data via the centroid approach or not including the masked data at all.

\begin{table}[tbp]
    \centering
    \footnotesize
    \begin{tabular}{c|c|c}
    & $\tilde{S}(s)$ & $p(s)$\\
     %& \multicolumn{2}{c}{$S(s)$} & %\multicolumn{2}{c}{$\text{logit}(p(s))$} \\
    \hline
    1a & 30.2 (16.3) &  19.0 (8.81) \\
    2a & 30.5 (17.5) & 20.2 (9.82)  \\
    3a & 33.6 (19.0) & 20.3 (9.73) \\
    1b & 45.4 (26.1) & 20.7 (9.79) \\
    2b & 43.8 (26.4) & 20.9 (9.66) \\
    3b & 45.8 (23.9) & 21.0 (9.73) \\
    4a & 46.2 (26.0) & 29.0 (14.0) \\
    5a & 43.6 (26.3) & 25.2 (13.2) \\
    6a & 38.5 (22.1) & 24.9 (12.4)\\
    4b & 65.4 (28.9) & 27.7 (12.4) \\
    5b & 85.7 (35.0) & 27.4 (13.5) \\
    6b & 56.2 (25.8) & 26.3 (12.7) 
    \end{tabular}
    \caption[MSE of predicted spatial and probability surfaces from the various models]{MSE (bias$^2$) of the probability surface from the various models on a 1km $\times$ 1km grids. All values have been multiplied by 100.}
    \label{tab:ulocs:mse}
\end{table}

% \begin{table}[tbp]
%     \centering
%     \footnotesize
%     \begin{tabular}{c|cc|cc}
%      & \multicolumn{2}{c}{$S(s)$} & \multicolumn{2}{c}{$\text{logit}(p(s))$} \\
%     Model & 1km & 5km & 1km & 5km\\
%     \hline
%     1a & 30.2 (16.3) & 30.2 (16.3) & 19.0 (8.81) & 19.0 (8.80) \\
%     2a & 30.5 (17.5) & 30.5 (17.5) & 20.2 (9.82) & 20.2 (9.80)  \\
%     3a & 33.6 (19.0) & 33.6 (19.0) & 20.3 (9.73) & 20.3 (9.70) \\
%     1b & 45.4 (26.1) & 45.4 (26.0) & 20.7 (9.79) & 20.8 (9.82) \\
%     2b & 43.8 (26.4) & 43.8 (26.3) & 20.9 (9.66) & 21.0 (9.66)\\
%     3b & 45.8 (23.9) & 45.7 (23.8) & 21.0 (9.73) & 21.1 (9.75) \\
%     4a & 46.2 (26.0) & 46.1 (26.0) & 29.0 (14.0) & 29.0 (14.0) \\
%     5a & 38.5 (22.1) & 38.5 (22.0) & 24.9 (12.4) & 24.9 (12.4)\\
%     4b & 65.4 (28.9) & 65.3 (28.8) & 27.7 (12.4) & 27.9 (12.5) \\
%     5b & 56.2 (25.8) & 56.1 (25.6) & 26.3 (12.7) & 26.4 (12.7)
%     \end{tabular}
%     \caption[MSE (bias$^2$) of predicted spatial and probability surfaces from the various models considered on 2 different grids]{MSE (bias$^2$) of the probability surface from the various models considered on 2 different grids. All values have been multiplied by 100.}
%     \label{tab:ulocs:mse}
% \end{table}

\begin{figure}[tbp]
    \centering
    \includegraphics[width=0.9\linewidth]{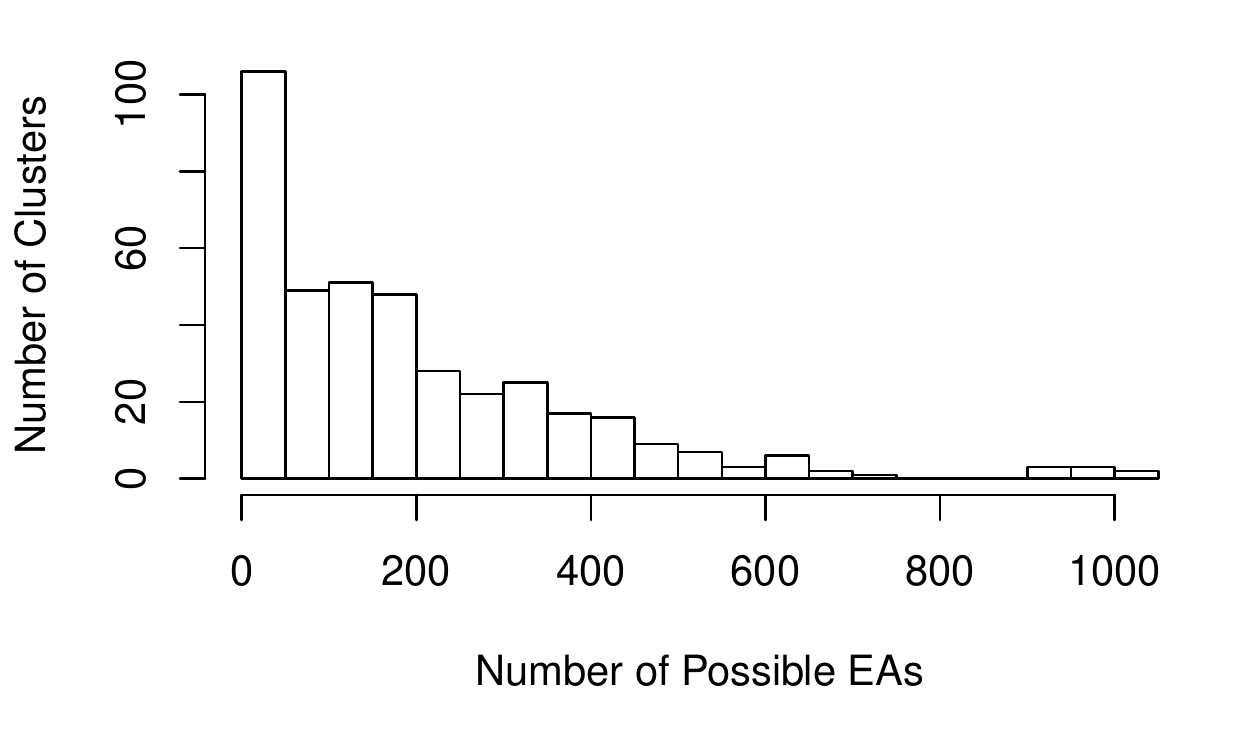}
    \caption{Histogram of potential disclosure risk.}
    \label{fig:histea}
\end{figure}

An important consideration is the disclosure risk, or ability to identify the enumeration area a particular set of data arose from. Exact identification in the jittering case would be possible if there is only 1 possible EA within 2km for urban coordinates or within 10km for rural coordinates. In our example, 1 cluster could be exactly identified with another 10 having only at most 5 possible EAs; see Figure \ref{fig:histea}.

 Another potential avenue for disclosure risk is if the posterior probability is significantly larger for one particular EA than for the other potential ones. To establish this, the posterior probability of the possible EAs is calculated and the largest and second largest are compared. We do this for scenario 3a. First, we note that for 6 clusters, there was 1 possible EA with posterior probability $> 0.95$, meaning that for those clusters disclosure risk is highly probable. Additionally, for 26 (105) clusters the most likely EA had a posterior probability that was more than 5 (2) times higher than the second most likely.

\begin{figure}[tbp]
    \centering
    \includegraphics[width=0.9\linewidth]{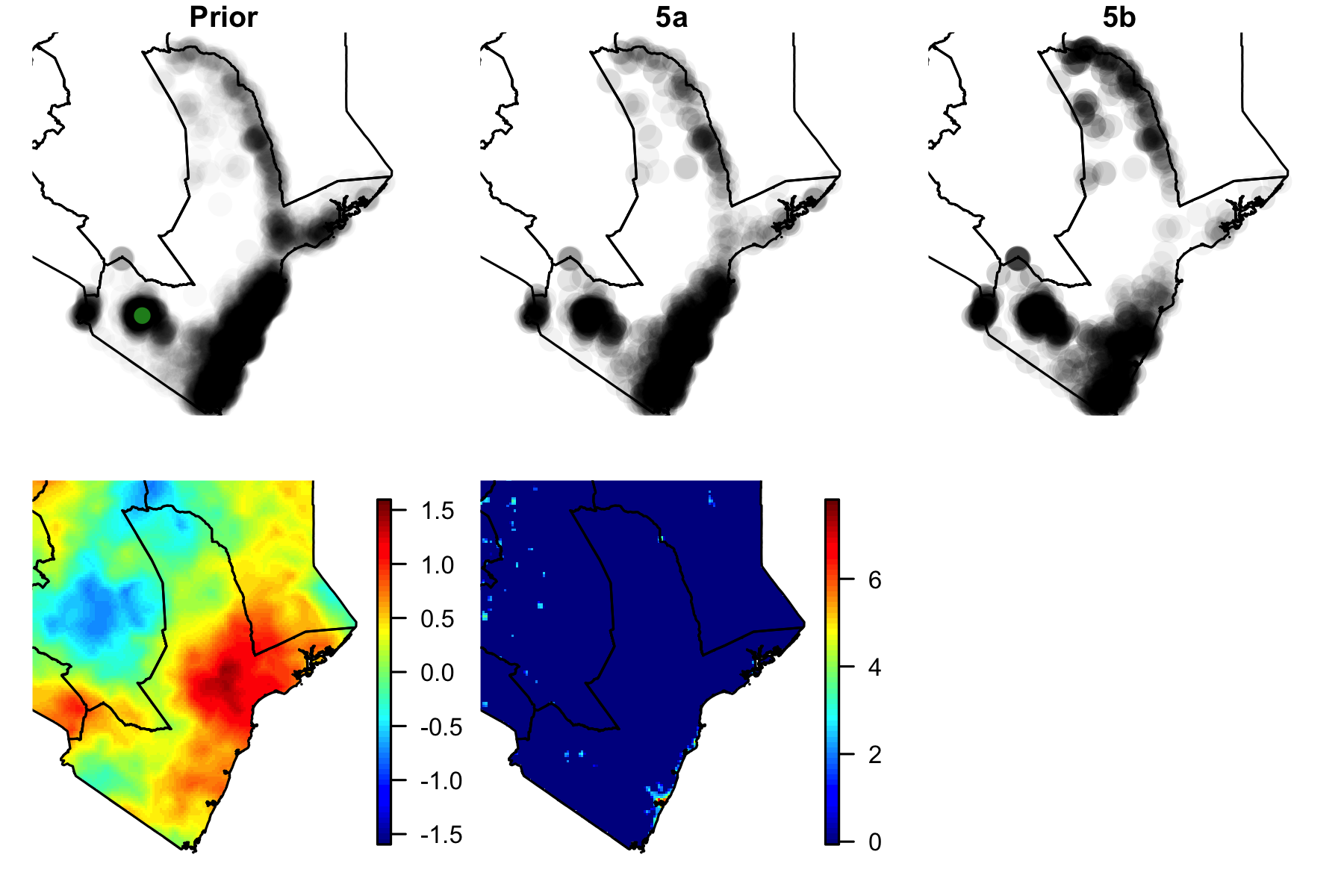}
    \caption[Posterior probability of EA location for masking scenario.]{Prior and posterior probability of EA location for masking scenario. Darker indicates higher probability. Green point is the true location of the cluster. Also shown is the latent spatial surface (bottom left) and light surface (bottom middle).}
    \label{fig:ulocs:locpost}
\end{figure}

This is less of a concern for the masking procedure as the number of possible EAs for each cluster range from 433 to 19,097 and the posterior probabilities were fairly uniform. Figure \ref{fig:ulocs:locpost} shows the prior and posterior probabilities for one cluster that was known to be from a rural EA from the Coast province with the outcome $y=5$ for 5a (no spatial covariate) and $y=2$ for 5b (spatial covariate). Noticeably, the posterior probability is lower than the prior probability in the central eastern region where the latent spatial surface is highest.

\section{Discussion}
\label{sec:disc}

In this paper, we propose an approach for incorporating data with missing or jittered GPS location information. We develop an ``INLA within MCMC'' approach, where we alternate between (1) updating the location of clusters by sampling from the full conditional posterior and (2) fitting conditional models using INLA and sampling from the approximated conditional posterior. In terms of computation time, it took about 52 hours to run 1,000 iterations for each scenario. The main computational burden comes from fitting the 1,000 \texttt{R-INLA} models.

We show that inference tends to be improved when the procedure that results in missing location information is taken into account through a simulation. Jittering of the coordinates did not have a significant impact on the results and one could argue that the more complicated DA procedure is not warranted. Further, there is a very real risk of identifying the true locations of (some of) the clusters, which is a privacy concern. From the DHS Terms of Use, users of geographic data ``agree to treat all data as confidential, and to make no effort to identify any individual, household, or enumeration area in the survey''
\nocite{dhsweb} 
(\url{https://dhsprogram.com/data/terms-of-use.cfm}). We illustrate that in our simulation this is possible if the potential sampling locations (i.e.,~the masterframe) are available.  Ideally, the jittering should provide a balance between accuracy in inference if the (incorrect) geographic coordinates are used and confidentiality in terms of not being able to uniquely identify the enumeration area that the cluster comes from. 

%Assuming that there should be at least two enumeration areas possible for a reported location to have come from, one approach would be to calculate the distance between all of the points and their closest point within each strata (e.g.,~administrative area and urban/rural). The minimum radius needed to guarantee two possible enumeration areas is double the maximum distance to the closest point \textbf{(JON: I'm not sure I agree with this statement anymore... do you understand it?)}. This would not necessarily guarantee posterior probabilities that are not very close to 1, but is a potential strategy for reducing some disclosure risk.
%Notably, this is less of a concern when the data is masked and this is the setting where we saw greatest improvements in inference.

% It should be noted that this is one simulation and thus the results are not necessarily indicative of what should happen. In the jittering scenario, we would expect the naive method to perform worse, especially since we used a covariate that varied quickly in space.

In our model formulation, access to a masterframe was assumed. However, in many cases a masterframe is not available; therefore, the possible cluster locations are unknown. In this case, one could be created as it was done for the simulation and assumed to be correct or the grid cells of a population density raster could be used as a surrogate, with the population density value of the grid cell being used in (\ref{eq:maskprior}).

Validity of the ``INLA within MCMC'' computational approach we employ relies on the validity of the INLA approximation. In our simulation, this approach seems to be accurate enough (in comparison to the models that could be fit solely in INLA) based on the posteriors obtained. To fully evaluate this approach, it would be best to compare results to only using MCMC. More practically, another option includes trying cruder approximations, i.e.,~the Gaussian and simplified Laplace approximations \citep{rue:etal:17} in the INLA step to see if the results seem stable. %, as suggested by \cite{rue:etal:09}.
Another strategy could be to refine the hyperparameter grid that is used in the approximation. One could also validate overall results by holding out data and then comparing results on a large administrative area level.

Future work involves expanding the scope of the simulation study to cases where the masterframe is not available, investigating the impact of other covariates that are smoother in space, altering the spatial range of the underlying process, and including different levels of masking. In the simulation, we supposed that the the available location information was the provincial level, but other geographies exist such as the county level. 
%Additionally, we plan to apply this method to survey data from Kenya.

\clearpage
%\section*{References}
%\bibliographystyle{natbib} 
%\bibliographystyle{plain} 
%\bibliography{/Users/jonno/Dropbox/BibFiles/spatepi}
\bibliography{spatepi-katie,spatepi}

%{\section{Supplementary Materials}
%\label{appendix:B}
%
%CODE}
\end{document}